\documentstyle[twocolumn,seceq]{jpsj}
\input epsf.tex
\def\ibid{{\it ibid}.\ }
\def\eg{{\it e.g.}\ }
\def\ie{i.e.\ }
\def\etal{{\it et\ al.\ }}

\newcommand{\lsim}
 {\ \raise.35ex\hbox{$<$}\kern-0.75em\lower.5ex\hbox{$\sim$}\ }
\newcommand{\gsim}
 {\ \raise.35ex\hbox{$>$}\kern-0.75em\lower.5ex\hbox{$\sim$}\ }
\def\journal #1#2#3#4{#1 {\bf #2} (#4) #3}

\def\PRB{Phys.\ Rev.\ B}
\def\PRL{Phys.\ Rev.\ Lett.}

\def\APNY{Ann.\ Phys.\ (New York)}

\def\CJP{Can.\ J.\ Phys.}

\def\PL{Phys.\ Lett.}
\def\PLA{Phys.\ Lett.\ A}

\def\JMP{J.\ Math.\ Phys.}
\def\CMP{Commun.\ Math.\ Phys.}

\def\JPA{J.\ Phys.\ A}
\def\JPC{J.\ Phys.\ C}
\def\JPCM{J.\ Phys.\ Cond.\ Mat.}
\def\JPCS{J.\ Phys.\ Chem.\ Solids}

\def\JPSJ{J.\ Phys.\ Soc.\ Jpn.}
\def\MPLB{Mod.\ Phys.\ Lett.\ B}

\def\runtitle{
Transition and Incommensurate Correlation in Spin-1/2 Heisenberg 
Chains
}
\def\runauthor{Shinji {\sc Watanabe} and Hisatoshi {\sc Yokoyama}}

\title
{
Transition from Haldane Phase to Spin Liquid and Incommensurate 
Correlation in Spin-1/2 Heisenberg Chains
}

\author
{
Shinji {\sc Watanabe}\footnote{E-mail: watanabe@cmpt01.phys.tohoku.ac.jp}
and Hisatoshi {\sc Yokoyama}
}

\inst
{
Department of Physics, Tohoku University, Aramaki Aoba, Aoba-ku,
Sendai 980-8578\\
}

\recdate
{
\today
}

\abst
{
The bulk properties and impurity effect in the one-dimensional $S=1/2$ 
Heisenberg model with dimerization and next-nearest-neighbor exchange 
are studied by the density matrix renormalization group and exact 
diagonalization methods. 
This model widely covers the $S=1$ Heisenberg model, two-leg ladders, 
bond-alternating chains and the $S=1/2$ Heisenberg model. 
Almost the whole parameter space belongs to the Haldane phase. 
In the non-dimerizing limit a transition to a kind of spin liquid 
takes place, characterized by the delocalized spin polarization. 
This transition causes the distinct impurity effects between a 
spin-Peierls compound CuGeO$_3$ and the Haldane systems. 
Experimental aspects of a bond-alternating compound $\rm CuNb_{2}O_{6}$ 
are also discussed. 
Furthermore, the incommensurate correlation arising in this model 
is extensively studied, and a phase diagram is constructed. 
}

\kword
{
spin gap, Haldane phase, spin-Peierls transition, incommensurate 
correlation, impurity effect, two-leg ladder, CuGeO$_3$, 
CuNb$_2$O$_6$, DMRG, exact diagonalization
}

\begin{document}
\sloppy
\maketitle

\section{Introduction}
Recent progress of experimental studies has shed light on
a variety of intriguing features in low-dimensional magnets.
Above all, spin gaps found in such materials arouse broad
interests with relation to the pseudo spin gap of the
high-$T_{\rm c}$ superconductors.
We list some classes of quasi one-dimensional (1D) spin-gap systems
with typical compounds:
The Haldane systems (NENP), bond-alternating systems (CuNb$_2$O$_6$),
spin-Peierls systems (CuGeO$_3$), two-leg ladder systems (SrCu$_2$O$_3$),
and so forth.
Although these compounds have similar spin-gap properties 
without any magnetic order in low temperatures, the behaviors
toward the elemental substitution are entirely different.\cite{Ajiro,Kojima}
A tiny amount of nonmagnetic or magnetic impurity (less than a 
few percent) breaks the spin gaps of CuGeO$_3$\cite{Hase,Manabe}
and SrCu$_2$O$_3$\cite{Azuma,Fujiwara}, and brings about 
antiferromagnetic (AF) long-range orders.
In contrast, some Haldane systems\cite{Ajiro,ItoKawae} and
CuNb$_2$O$_6$\cite{Nishikawa} keep their spin-gap behaviors
even when substitutions as much as 20-40\% are made, and
no signs of magnetic orders are observed.
\par

As the origin of this difference, multiple factors can be considered:
[1] Difference in the nature of spin states within chains, 
[2] higher-dimensional (interchain) magnetic couplings, 
\cite{Sakai,TroyerZU} 
[3] couplings with lattice degrees of freedom,\cite{Khomskii} and
[4] other factors, for example, disorder induced by the impurities, 
etc.
Among these, we notice the factor [1], allowing for 
the following facts. 
Various experimental results for the above compounds are 
systematically and quantitatively understood within pure 1D spin 
models.\cite{RD-Castilla,YS,Klumper,Nishikawa,KodamaINS} 
Furthermore, in spite of the similar interchain couplings 
(about 10\% of the intrachain coupling\cite{Nishi,KodamaINS}), 
CuGeO$_3$ and CuNb$_2$O$_6$ exhibit distinct behaviors toward 
the impurity, as mentioned. 
\par

\begin{figure}
\begin{center}
\epsfxsize=8cm \epsfbox{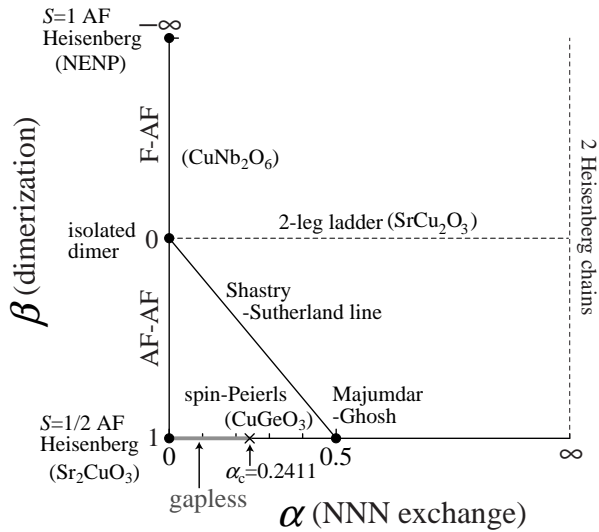}
\end{center}
\caption{
Schematic diagram of the parameter space for the Hamiltonian 
eq.~(\ref{eq:hamil}), spanned by two dimensionless parameters: 
$\alpha$ is of the next-nearest-neighbor exchange and $\beta$ 
of the bond alternation. 
}
\label{fig:ab}
\end{figure}

The main purpose of this paper is to investigate the cause 
of the above impurity effect on the basis of 1D chains. 
Using the density matrix renormalization group (DMRG), exact 
diagonalization (ED) methods and the perturbation theory, 
we study the properties within 1D precisely, thereby discuss 
the interchain coupling and the lattice effect. 
In order to argue in a unified viewpoint, we consider a model 
which continuously connects all the above materials in its 
parameter space: The 1D spin-1/2 Heisenberg model with 
next-nearest-neighbor (NNN) coupling and bond 
alternation,\cite{Brehmer1,Brehmer2,Naka1}
\begin{eqnarray}
\label{eq:hamil}
{\cal H}
&=&J\left[\;\sum_{i=1}^{L/2}\,\Bigl({\mib S}_{2i-1}\cdot{\mib S}_{2i}
          + \beta\,{\mib S}_{2i} \cdot {\mib S}_{2i+1}\Bigr)\right.
\qquad\quad \nonumber \\
&&\qquad\qquad\qquad\qquad
  +\left.\alpha \sum_{i=1}^L\,{\mib S}_{i} \cdot {\mib S}_{i+2}\;\right],
\end{eqnarray}
where $J>0$, and $\alpha$ and $\beta$ are dimensionless parameters 
controlling the coupling strength (see Fig.~\ref{fig:ab}). 
\par

First, we study the bulk properties of the ground state and 
low-lying excited states, to grasp the global feature the Hamiltonian 
possesses in its parameter space. 
Next, we study the properties of finite systems, namely segments 
separated by the impurities, paying attention to the edge condition. 
In particular, the local spin polarization is important to 
consider magnetic long-range orders. 
The key point of the present impurity issue is that in the transition 
area from the Haldane phase to the spin liquid phase the local spin 
polarization highly depends on the system size.
\par

Another fascinating aspect of this model is the incommensurate
(IC) correlation.
So far, except for the case without dimerization 
($\beta=1$)\cite{Tonegawa,Bursill1,WhiteA} few properties of 
the IC correlation have been elucidated.\cite{Chitra} 
To complete the phase diagram of the IC correlation in the 
$\alpha$-$\beta$ plane is another important purpose of this 
paper.
\par
%
This paper is organized as follows. 
In \S2 we briefly review the previous studies on the model 
eq.\ (1.1). 
In \S3 the bulk properties of several quantities are studied 
widely in the $\alpha$-$\beta$ plane. 
In \S4 the incommensurate correlation is studied. 
In \S5 we concentrate on the regime of spin-Peierls compounds 
($\beta\sim 1$), and consider the impurity effect in $\rm CuGeO_{3}$.
In \S6 the regime of Haldane phase is discussed with CuNb$_2$O$_6$ 
in mind.
Finally, we summarize the main results in \S7. 
In Appendix A, the relations between the two representations
of the Hamiltonian, eq.~(\ref{eq:hamil}) and eq.~(\ref{eq:hamil2}), 
are given. 
In Appendix B we supplement the explanation in \S3.1 on the 
anomalous behaviors of the spin gap. 
In Appendix C we study the low-lying level structure for 
different boundary conditions, and show that the transition 
from the Haldane phase to the spin liquid occurs at $\beta=1$. 
In Appendix D we study the dynamical properties of the bond-alternating
chains, in addition to various properties of monoclinic 
$\rm CuNb_{2}O_{6}$. 
\par

\section{Model and Method}
\begin{figure}
\begin{center}
\epsfxsize=8cm \epsfbox{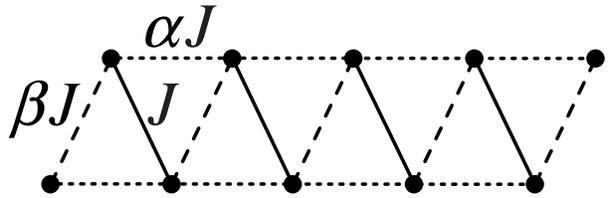}
\end{center}
\caption{
Connectivity of the Hamiltonian eq.~(\ref{eq:hamil}). 
}
\label{fig:ladder}
\end{figure}

The Hamiltonian eq.~(\ref{eq:hamil}) is represented as a two-leg 
ladder with zigzag rungs as shown in Fig.~\ref{fig:ladder}.
This model includes various classes of 1D spin systems as described 
below. 
The parameter of bond alternation $\beta$ is restricted to the
range $-\infty\le\beta\le 1$, by taking account of the translational 
symmetry. 
For the open boundary condition (OBC), however, it is quite 
important whether the first and last bonds have the value $J$ 
or $\beta J$.
This point will be discussed later. 
As for the parameter of the NNN exchange $\alpha$, we restrict 
the range to $0\le\alpha\le\infty$ in this paper.
\par
There are some special regions in the $\alpha$-$\beta$ plane 
as listed below (See also Fig.~\ref{fig:ab}):                      
\begin{enumerate}
\item[ [1]$\!\!\!$]
$(\alpha,\beta)=(0,1)$: The $S=1/2$ AF Heisenberg model. 
\par
\item[[2]$\!\!\!$]
$\beta=-\infty$: The $S=1$ AF Heisenberg model.          
\item[[3]$\!\!\!$]
$\beta=2\alpha$ ($0\le\alpha\le 0.5$): The so-called Shastry-Sutherland 
(SS) line.\cite{SS} 
In particular, the two ends $(0,0)$ and $(0.5,1)$ are (a) the 
isolated dimer point, (b) the Majumdar-Ghosh (MG) point,\cite{MG} 
respectively. 
\item[[4]$\!\!\!$]
$\beta=0$: The two-leg ladder.                    
The isotropic ladder ($J_{\rm leg}/J_{\rm rung}=\alpha=1$) is 
represented by $(1,0)$. 
\item[[5]$\!\!\!$] $\alpha=\infty$: The model is decoupled to 
two $S=1/2$ AF Heisenberg chains with the doubled unit cell. 
\end{enumerate}
\par

We start with the relation to the Haldane issue.
On the $\beta$ axis ($\alpha=0$), the Hamiltonian represents 
the AF-ferromagnetic (F) bond-alternating chains for $\beta<0$. 
In the limit of $\beta\rightarrow -\infty$ (as the Hund coupling), 
$\cal H$ is exactly reduced to the $S=1$ Heisenberg model with 
a numeric factor 1/4. 
In this viewpoint Hida studied the properties for finite values of
$\beta$ ($\alpha=0$), using ED.\cite{Hida}
He found that the properties of the Haldane phase continue even 
to the AF-AF alternating region ($\beta>0$). 
Later, by extending the non-local unitary transformation used
for the $S=1$ systems,\cite{KennedyT} Kohmoto and Tasaki\cite{KohmotoT}
and Takada\cite{Takada} showed exactly that the hidden $Z_2\times Z_2$
symmetry is completely broken for the case [3(a)], and approximately
that the Haldane phase continues to the limit [1].\cite{HidaT}
\par

On the other hand, there are many researches in view of the 
analogy between the Haldane systems and ladders. 
For example, White showed that the isotropic ladder point [4] 
belongs to the same phase with the Haldane system, using 
DMRG.~\cite{WhiteHL}
Recently using a variational matrix product method, Brehmer \etal
studied the ground-state properties\cite{Brehmer1} and low-energy
excitations\cite{Brehmer2} of eq.~(\ref{eq:hamil}), and concluded
that their trial function smoothly connects the AKLT model\cite{AKLT} 
and the MG model [3(b)]. 
Thus, a wide range of the $\alpha$-$\beta$ space seems covered
by the Haldane phase.
In this phase the following conditions have to be satisfied:
\begin{itemize}
\item
A unique singlet ground state with a finite singlet-triplet 
excitation gap (Haldane gap) for the systems of $L=\infty$ or 
for the periodic boundary condition (PBC). 
\item
Fourfold (pseudo-)degeneracy appears under the Haldane gap 
for the finite open chains due to the two localized $S=1/2$ 
spins at both edges.\cite{Kennedy}
\item
A hidden topological (string) long-range order (LRO) exists.\cite{Nijs}
\end{itemize}
\par

At the MG point [3(b)], the ground state for the PBC spontaneously 
breaks the translational symmetry and has twofold degeneracy: 
\begin{eqnarray}
\label{eq:psi1}
      \psi_{1} &=& [1,2][3,4]\cdots[L-1,L],  \\
      \psi_{2} &=& [2,3][4,5]\cdots[L,1],
\end{eqnarray}
where $[i,j]=\alpha(i)\beta(j)-\beta(i)\alpha(j)$.
On the other hand, on the SS line $\beta$=2$\alpha$ ($0\le\alpha<0.5$), 
$\psi_{1}$ whose singlet pairs sit on the bonds with the larger 
exchange $J$ becomes the unique ground state, and $\psi_{2}$ is 
no longer an eigenstate. 
\par

On the $\alpha$ axis ($\beta=1$),\cite{Tonegawa} the excitation 
is gapless for $0\le\alpha\le \alpha_{\rm c}$. 
Note that the gapless regime in the $\alpha$-$\beta$ plane 
is limited to this segment and $\alpha=\infty$. 
The critical point $\alpha_{\rm c}=0.241167$\cite{OkamotoN} 
is connected to the Haldane-Shastry model,\cite{HS} in the sense 
that the conformal invariance is realized.
For $\alpha>\alpha_{\rm c}$, the ground state is twofold
degenerate for $L\rightarrow\infty$, in accordance with the 
Lieb-Shultz-Mattis theorem.\cite{LiebSM}
\par

The IC correlation arises in the $\beta<2\alpha$ side of the SS 
line.\cite{Tonegawa,Chitra,Bursill1,WhiteA}
However, details including the boundary position have been still 
obscure. 
As far as the $\alpha$ axis ($\beta=1$) is concerned, it is 
known that an IC wave number which maximizes $S(q)$\cite{noteq} 
appears for $0.52063\le\alpha\le\infty$~\cite{Bursill1,WhiteA}. 
\par

Here, we comment on another and more popular representation 
in the spin-Peierls regime ($\beta\lsim 1$): 
\begin{equation}
\label{eq:hamil2}
 {\cal H} = \tilde J\sum_{i}\Bigl\{[1-(-1)^{i}\delta]
              {\mib S}_{i} \cdot {\mib S}_{i+1}
          +\tilde\alpha{\mib S}_{i} \cdot {\mib S}_{i+2}\Bigr\}. 
\end{equation}
The Hamiltonians eqs.~(\ref{eq:hamil}) and (\ref{eq:hamil2}) 
are identical except that the space of eq.~(\ref{eq:hamil2})
is limited to $1\ge\beta\ge -1$ in the $\alpha$-$\beta$ plane.
We use each representation in its suitable situation. 
The relations between the two representations are summarized
in Appendix A. 
\par

Lastly, we describe numerical methods used.
The DMRG method for zero temperature~\cite{WhiteM} is a powerful 
method to calculate the ground-state and the low-energy 
properties accurately for low-dimensional systems. 
In this paper we adopt the finite size algorithm with OBC.
The number of states kept for each block is up to $m=200$.
The system size is usually $L\sim 200$, and is increased
up to 400 according to necessity. 
The typical truncation error is $O(10^{-12})$ for the region
away from the $\alpha$ axis ($\beta=1$).
But it becomes $O(10^{-8})$ 
as $\beta$ approaches 1, where the correlation
length diverges, and also in the incommensurate area.
We show error bars when accuracy is required.
\par

The ED (Lanczos and Householder) methods are used for supplementary
calculations to DMRG, and to study dynamical and thermodynamic
properties with a recursion procedure for the former.
We check system-size dependence and make extrapolations from
the data up to $L=26$ for the former properties and up to $L=18$
for the latter.
\par

\section{Global Features of the $\alpha$-$\beta$ Plane}
In this section, we study the bulk properties of several 
quantities widely in the $\alpha$-$\beta$ space. 
In \S3.1 we consider the spin gap. 
The anomaly in the spin gap is a sign of the incommensurate 
correlation treated in \S4.
In \S3.2 two-point correlation function is discussed.
In \S3.3 the order parameters of dimerization and of hidden
topological (string) order are studied.
We refer to the phase transition from the Haldane phase to 
a spin liquid at $\beta=1$.
\par

\subsection{Spin gap}\label{chap:sgap}
Before discussing the results, we need to remark on the edge effect 
in the DMRG calculations of the gap.
As is known, an open $S=1$ chain has free $S=1/2$ spins near 
both chain ends in the Haldane phase. 
These spins yield pseudo fourfold degenerate states (one singlet
and three triplets) beneath the bulk Haldane gap. 
Therefore, in estimating the bulk gap by DMRG,\cite{WhiteH} extra 
$S=1/2$ spin had to be attached to both edges artificially to 
cancel the free spins. 
In the present case we take advantage of a similar but natural
cancellation, by imposing the condition on the system that the 
edge bonds should be $J$ (the stronger one). 
For the bulk quantities, we use this boundary condition. 
We will discuss the effect of edge bond in Appendix C.2. 
\par

Figure~\ref{fig:gapb} shows the $\beta$ dependence of the spin gap 
$\Delta$ for some values of $\alpha$ obtained by DMRG. 
The spin gap is obtained by $\Delta(L)=E(1;L)-E(0;L)$, 
where $E(S^{z};L)$ is the lowest energy with total $S^{z}$ and 
system size $L$. 
We extrapolate $\Delta(L)$ to the thermodynamic limit by using 
the polynomial fit:
$$
\Delta(L)=\Delta(\infty)+\sum_{n=1}\frac{c_{n}}{L^{n}}, 
$$
with $n\le 8$ for $L=12,16,24,32,48,64,100,200,400$. 
\par

For $\alpha=0$, the result is consistent with those previously 
estimated by ED.\cite{Hida}
For $\beta \to -\infty$ the gap is smoothly reduced to
$\Delta_{\rm H}/4J$, where $\Delta_{\rm H}=0.41050(2)$~\cite{WhiteH}
is so-called Haldane gap in unit of the exchange coupling
for the $S=1$ AF Heisenberg chain. 
\par
\begin{figure}
\begin{center}
\epsfxsize=8cm \epsfbox{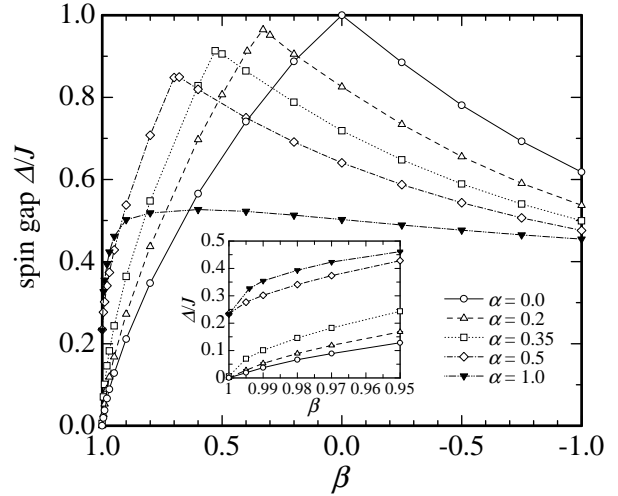}
\end{center}
\caption{$\beta$-dependence of the spin gap for some values of
$\alpha$ estimated by the DMRG method. 
The values are obtained by the extrapolation of the finite 
systems of $L=12$-$400$. 
The inset shows the enlargement for $0.95 \le\beta \le 1$. 
}
\label{fig:gapb}
\end{figure}
%
At the isolated dimer point the gap becomes the maximum $J$ in the 
$\alpha$-$\beta$ plane.
There, the model is reduced to the two-spin problem, so that 
the energy cost to break up one singlet into triplet is $J$. 
At this point $\Delta$ forms a cusp as seen in Fig.~\ref{fig:gapb}. 
In the following, we study this anomaly in connection with the 
incommensurate correlation treated in \S4. 
For finite but small values of $\alpha$ and $\beta$, this anomaly 
can be traced by the second-order perturbation theory. 
The energy per bond for the ground state (singlet) and the lowest 
triplet branches are obtained respectively as, 
\begin{eqnarray}
\frac{E_{\rm s}}{J} &=& -\frac{3}{4}-\frac{3}{16}(\beta-2\alpha)^{2}. 
\label{eq:Esdimer} \\
\frac{E_{\rm t}(q)}{J} &=& \frac{1}{4}-\frac{1}{4}(\beta^{2}+4\alpha^{2}) 
\label{eq:Etdimer}          \\
           & & -\left[ \frac{1}{2}(\beta-2\alpha)
             +\frac{1}{4}(\beta^{2}-4\alpha^{2})\right]\cos(2q),
\nonumber 
\end{eqnarray}
where $q=n\pi/N$ ($-N/2<n \le N/2;\ n={\rm integer}$) and $N=L/2$ 
(the number of dimers) with the lattice constant being unity. 
Note that for $\beta=2\alpha$ cancellation occurs between 
$\alpha$ and $\beta$.
For the ground state, $E_{\rm s}$ leads to the constant exact 
value of the SS line~\cite{SS}.
Meanwhile, for $E_{\rm t}(q)$ the hopping matrix element in 
eq.~(\ref{eq:Etdimer}) vanishes, so that the singlet-triplet 
excitation forms a flat blanch. 
When a parameter ($\alpha$ or $\beta$) is varied and crosses 
this line, the wave number minimizing $E_{\rm t}(q)$ switches 
from $q=\pi/2$ to $0$ $(\pi)$ discontinuously, due to the hopping 
term. 
Thus, the above dispersionless point corresponds to the cusp of 
the spin gap. 
On the other hand, the second term in eq.~(\ref{eq:Etdimer}) 
survives even for $\beta=2\alpha$; the spin gap decreases from 
$J$ even along the cusp line. 
\par

When ($\alpha$,$\beta$) goes away from (0,0), the cusp still 
remains as seen in Fig.~\ref{fig:gapb}. 
In the $\alpha$-$\beta$ plane, the cusp point deviates to 
the side of $\beta <2\alpha$ with increasing $\alpha$, as 
shown in Fig.~\ref{fig:abplane}. 
This singular point is related to the incommensurate correlation, 
and will be taken up again in \S4. 
We have confirmed by ED that the properties found by the above 
perturbation theory are preserved for $\alpha \lsim 0.5$. 
On the other hand, for $\alpha \gsim 0.5$ with increasing $\beta$, 
the bottom of the triplet blanches does not switch directly 
from $\pi/2$ to $0$ ($\pi$), but IC values of $q$ intervene. 
Thus, the cusp disappears. 
Actually, for $\alpha=1$ the gap changes smoothly from the Haldane 
limit ($\beta=-\infty$) to $\beta \sim 1$ through the isotropic 
ladder point ($\beta$=0), as depicted in Fig.~\ref{fig:gapb}.  
The supplement of this issue is given in Appendix~B. 
\par

The inset of Fig.~\ref{fig:gapb} shows the enlargement for 
$0.95 \le \beta \le 1$. 
The $\delta$-dependence\cite{sP} of the gap has been discussed 
in relation to the spin-Peierls transition.\cite{BonnerB,YS} 
The gap drops steeply near $\beta=1$ even for the spontaneously 
gapped regime ($\alpha > \alpha_{\rm c}$), as seen for 
$\alpha=0.35$ and $0.5$. 
\par

\subsection{Spin-spin correlation}
In this subsection, we discuss the spin-spin correlation function
$S_{ij}=\langle S^{z}_{i}S^{z}_{j}\rangle$ in the ground state.
Using the system of $L=280$ typically, we calculate $S_{ij}$ 
in the central part of the system.
The 40 sites near the open edges are not used to eliminate edge 
effects. 
\par

\begin{figure}
\begin{center}
\epsfxsize=8cm \epsfbox{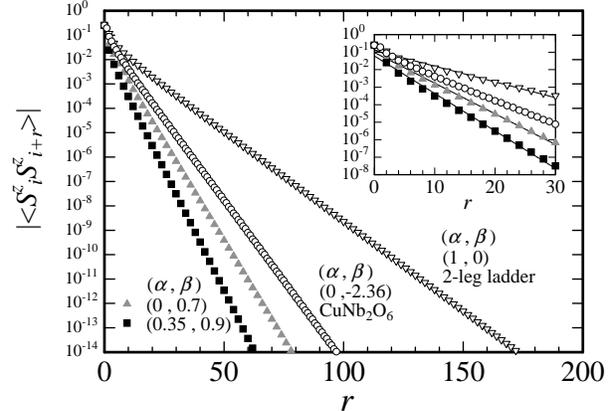}
\end{center}
\caption{
Spin-spin correlation function versus $r$ for some typical cases 
of exponential decay. 
Since the fitted line for odd $r$ shifts a little from that for 
even $r$ for $\beta>0$, only the data for even $r$ are shown 
except for $(0,-2.36)$. 
The inset is the magnification of the small-$r$ regime with fitted 
lines.
}
\label{fig:fit1}
\end{figure}

In the gapped regime, where $S_{ij}$ decays exponentially for
$|i-j|\to\infty$,\cite{note32} we assume a simple form:
\begin{eqnarray}
\label{eq:SSexp}
\langle S^{z}_{i}S^{z}_{i+r}\rangle \propto
\exp\left(-\frac{\xi}{r}\right). 
\end{eqnarray}
Incidentally, we do not adopt the 2D Ornstein-Zernike type
used for the $S=1$ chain,\cite{NomuraS1} because it is not
certain whether it is adequate also in the present model. 
The above fit successfully works for most cases except for 
$\beta\sim1$, where the algebraic decay is essential. 
In Fig.~\ref{fig:fit1} typical cases of the exponential decay 
are shown: An AF-AF and a F-AF bond-alternating systems and 
the isotropic ladder, etc. 
The inset shows that the ranges of ill fit by eq.~(\ref{eq:SSexp}) 
are limited to several lattice spacings. 
The correlation length $\xi$ is obtained from this fit.
\par

\begin{figure}
\begin{center}
\epsfxsize=8cm \epsfbox{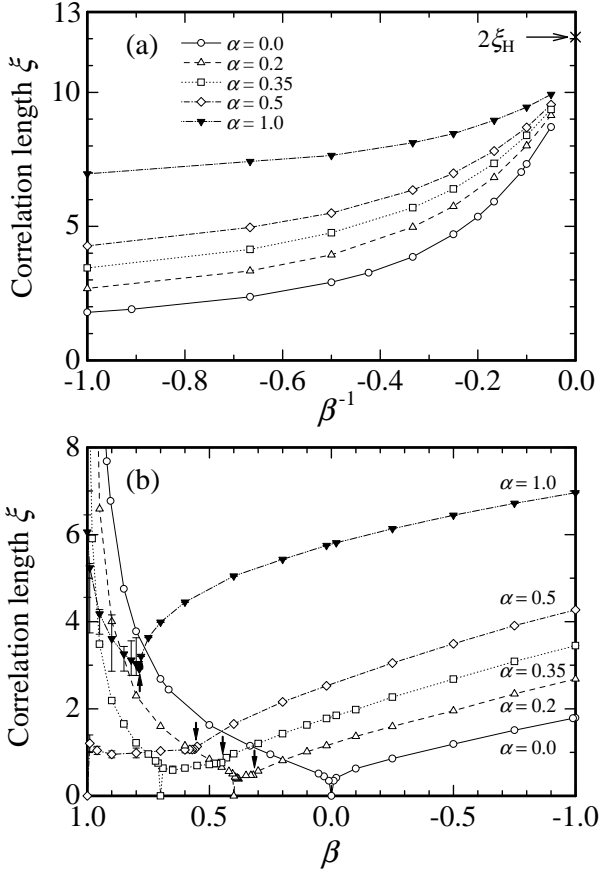}
\end{center}
\caption{
Correlation length for some values of $\alpha$ as a function 
of (a) $1/\beta$ for $-\infty<\beta\le-1$, and of (b) $\beta$ 
for $-1\le\beta<1$.
In (a) cross symbol at $\beta^{-1}=0$ indicates twice the 
correlation length of the $S=1$ AF Heisenberg model.~\cite{WhiteH} 
In (b) arrows stand for the commensurate-incommensurate boundary 
observed in $S_{ij}$. 
$\xi$ vanishes on the Shastry-Sutherland line ($\beta=2\alpha$).
}
\label{fig:crr2}
\end{figure}

%
In Figs.~\ref{fig:crr2}(a) and \ref{fig:crr2}(b) we plot $\xi$
for $-1\ge\beta>-\infty$ and  $1\ge\beta\ge-1$, respectively.
In the Haldane limit ($\beta \to-\infty$), $\xi$ is 
reduced to $2\xi_{\rm H}$ [cross symbol in Fig.~\ref{fig:crr2}(a)],
where $\xi_{\rm H}=6.03(1)$~\cite{WhiteH} is the correlation
length of the $S=1$ AF Heisenberg chain. 
The factor 2 arises due to the doubled unit cell. 
As $\alpha$ increases, $\xi$ approaches the value in the Haldane 
limit faster. 
This is because $\alpha$ does not work as frustration for 
$\beta <0$, but plays a cooperative role to stabilize 
the $\uparrow\uparrow\downarrow\downarrow$ structure. 
\par

As shown in Fig.~\ref{fig:crr2}(b) $\xi$ vanishes
on the SS line ($\beta=2\alpha$), where $S_{ij}$ vanishes
for $|i-j|\ge 2$.\cite{MG} 
Comparing with Fig.~\ref{fig:gapb}, we find the minimum point
of $\xi$ does not coincide with the maximum point of the spin gap, 
except for the isolated dimer point. 
In the upper side of the SS line ($\beta<2\alpha$), the region 
exists where $|S_{ij}|$ weakly oscillates with an IC period\cite{Chitra}. 
We indicate the other commensurate-IC boundary---so far unknown---by 
arrows in Fig.~\ref{fig:crr2}(b).
In this regime, the fit by eq.~(\ref{eq:SSexp}) becomes less 
accurate with increasing $\beta$ because of the frequent 
oscillation; possible errors thereof are shown by bars. 
We will discuss the IC correlation in \S4.
\par

\begin{figure}
\begin{center}
\epsfxsize=8cm \epsfbox{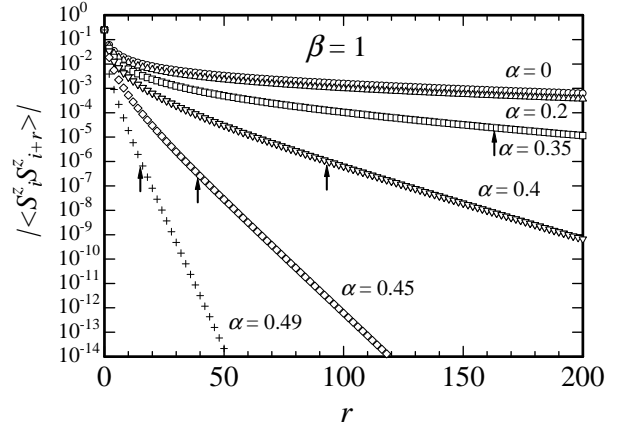}
\end{center}
\caption{
Spin-spin correlation function $\langle S^{z}_{i}S^{z}_{i+r}\rangle$ 
as a function of $r$ for some values of $\alpha$.
$\beta=1$. 
The arrow indicates $\eta$, which roughly divides $S_{ij}$ into 
two parts, of exponential decay ($r>\eta$) and of semi-algebraic 
decay ($r<\eta$). 
The data for even $r$ are shown for clarity. 
}
\label{fig:crr1}
\end{figure}

%
Now, let us consider the vicinity of the $\alpha$ axis.
In Fig.~\ref{fig:crr1} we show $S_{ij}$ for $\beta=1$. 
For the gapless regime $S_{ij}$ shows algebraic decay (with some 
logarithmic correction\cite{AffleckS}), as seen for $\alpha=0$ 
and $0.2$. 
An important fact is that even in the gapped regime 
($\alpha > \alpha_{\rm c}$), $S_{ij}$ shows a similar 
non-exponential decay for some distance. 
Of course, in a sufficiently long distance $S_{ij}$ decays 
exponentially. 
Therefore, we try to divide $S_{ij}$ into two parts. 
The arrows in Fig.~\ref{fig:crr1} show the minimal $r(\equiv\eta)$ 
which satisfies the condition that the error in the exponential fit 
for $r>\eta$ exceeds $10^{-3}$ (this value is for convenience). 
Thereby, we roughly estimate the boundary of the decaying behavior. 
For $\alpha<\alpha_{\rm c}$, $\eta=\infty$; even for 
$\alpha>\alpha_{\rm c}$, $\eta$ is still considerably large, 
\eg $\eta=163$ for $\alpha=0.35$. 
In the vicinity of the MG point ($\xi=0$) $\eta$ decreases 
abruptly. 
\par

This tendency still remains in the vicinity of the $\alpha$ axis 
($\beta\sim 1$). 
There, the short-range correlation is controlled by the 
semi-algebraic decay, and poorly fitted by eq.~(\ref{eq:SSexp}). 
Therefore, the values of $\xi$ plotted in Fig.~\ref{fig:crr2}(b) 
have been estimated from $S_{ij}$'s with sufficient large $|i-j|$. 
Such enhancement of the short-range spin correlation near the 
$\alpha$ axis is closely connected to the impurity effect in 
spin-Peierls systems. 
This will be discussed in \S5. 
\par
%
\subsection{Order parameters}\label{chap:order}
\begin{figure}
\begin{center}
\epsfxsize=8cm \epsfbox{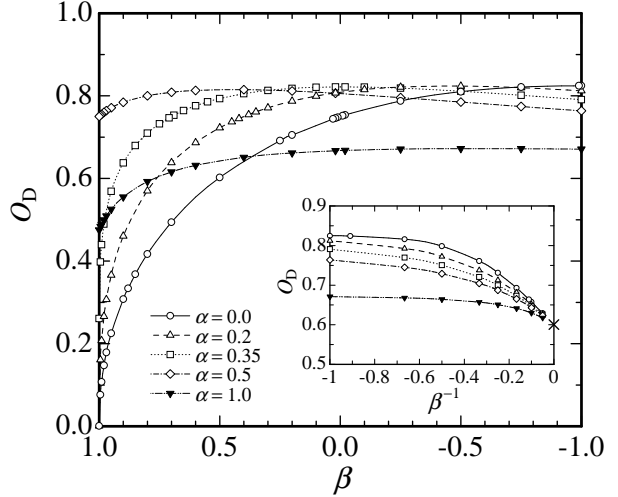}
\end{center}
\caption{
Dimer order parameter as a function of $\beta$ ($1\ge\beta\ge -1$)
for some values of $\alpha$.
Inset shows the same quantity for $\beta\le -1$ as a function of
inverse $\beta$.
The cross at $\beta^{-1}=0$ indicates the value for the $S=1$ 
Heisenberg chain: $(1-E_{\rm H})/4$, where 
$E_{\rm H}=-1.40148440390$.\cite{WhiteH}
On the SS line ($\beta=2\alpha$) $O_{\rm D}=3/4$.
Extrapolation is carried out from the DMRG data of $L=12$-$400$. 
}
\label{fig:SSdimer}
\end{figure}

%
We begin with an order parameter for dimerization defined by
\begin{eqnarray}
\label{eq:SSdimer}
O_{\rm D}=
\left|\langle{\mib S}_{2i-1} \cdot {\mib S}_{2i}\rangle-
\langle{\mib S}_{2i} \cdot {\mib S}_{2i+1}\rangle\right|. 
\end{eqnarray}
Figure~\ref{fig:SSdimer} shows $O_{\rm D}$ versus $\beta$ 
for some values of $\alpha$. 
\par 

First, we look at the cases for $\alpha=0$. 
As $\beta$ decreases from 1, the dimerization abruptly becomes 
larger and continues increasing until $\beta=-1$, where $O_{\rm D}$ 
has the maximum value $0.8245165$. 
Since $O_{\rm D}=3/4$ for the isolated dimer point, moderate 
magnitude of ferromagnetic exchange between singlets further 
promotes the dimerization.\cite{WhiteA} 
Beyond $\beta=-1$ $O_{\rm D}$ monotonically decreases and tend 
to the value $O_{\rm D}$=0.600371010 estimated for the $S=1$ AF 
Heisenberg chain, for $\beta\to -\infty$.
\par

%
Next, for $0<\alpha<\alpha_{\rm c}$ there remain the absence of 
dimerization for $\beta=1$ and the abrupt increase for $\beta\sim 1$. 
On the other hand, for $\alpha>\alpha_{\rm c}$, $O_{\rm D}$ shows 
a finite value even for $\beta=1$ due to the spontaneous dimerization. 
With increasing $\alpha$ the maximum point of $O_{\rm D}$ 
shifts in the positive direction of $\beta$, although the maximum 
value is almost unchanged. 
This aspect is understood by the cooperative roles of $\alpha$ 
and $-\beta$, as mentioned in \S3.2. 
In contrast to the case of $\beta =1$,\cite{WhiteA} there is 
no direct correspondence between the spin gap and $O_{\rm D}$. 
In Fig.~\ref{fig:gapb} the spin gap shows the cusp, while $O_{\rm D}$ 
does not show any singular behavior, and smoothly connects the 
Haldane limit ($\beta=-\infty$) and the limit of $\beta\to 1$.
\par

Now, we turn to the string order parameter $O_{\rm S}$.
The string correlation function to measure the topological
LRO in the Haldane phase for $S=1$ systems\cite{Nijs}
was extended to the $S=1/2$ bond-alternating chain as,\cite{HidaT}
\begin{eqnarray}
\label{eq:Ostr}
O_{\rm S}(i-j)=&& \nonumber \\
-\Bigl\langle (S^{z}_{2i-1}+S^{z}_{2i})&&
\prod_{\ell=2i+1}^{2j-2}\!\!\!\exp\Bigl(i \pi S^{z}_{\ell}\Bigr)
\,(S^{z}_{2j-1}+S^{z}_{2j})\Bigr\rangle.\qquad  
\end{eqnarray}
In this expression, ferromagnetic (weaker AF) bonds are
assumed between the sites $2i-1$ and $2i$ and between ones $2j-1$
and $2j$. 
Among others, this definition is the most natural extension to 
the present system.
\par

\begin{figure}
\begin{center}
\epsfxsize=8cm \epsfbox{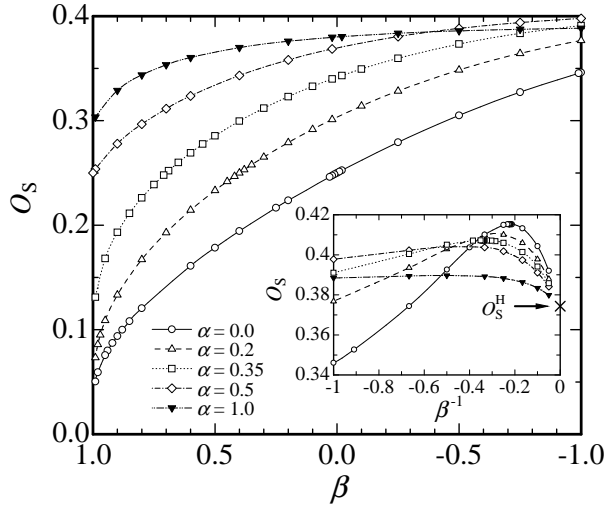}
\end{center}
\caption{
String order parameter as a function of $\beta$ ($1>\beta\ge -1$) 
for some values of $\alpha$. 
The inset shows the same quantity for $-1>\beta\ge -\infty$. 
The cross at $\beta^{-1}=0$ indicates
$O_{\rm S}^{\rm H}(\infty)=0.374325096(2)$ for the $S=1$
Heisenberg model.~\cite{WhiteH}
On the SS line $O_{\rm S}=1/4$.
}
\label{fig:Ostr}
\end{figure}

In Fig.~\ref{fig:Ostr} shown is $O_{\rm S}=\lim_{r\to\infty}O_{\rm S}(r)$ 
estimated by DMRG.
As has pointed out for the $S=1$ case,\cite{WhiteH} it is technically
important to measure $O_{\rm S}(r)$ at a sufficient distance from 
the open edge, not to mention large $|r|$.
Actually, for the cases of $\beta\sim 1$ and small $\alpha$,
the required distance exceeds our chain length ($L=400$).
This technical difficulty prevents us from concluding decisively
that $O_{\rm S}$ vanishes as $\beta\to 1$ for $\alpha<\alpha_{\rm c}$.
We believe it occurs, however, taking account of the singular 
behaviors near $\beta=1$.
On the other hand, for $\alpha>\alpha_{\rm c}$, $O_{\rm S}$ has a
finite value; for the MG point, $O_{\rm S}=1/4$.
\par

When $\beta$ decreases from $1$, $O_{\rm S}$ increases at first,
has a maximum at a finite value of $\beta$, and then tends to 
the value for the $S=1$ case. 
Here again, the value for a larger $\alpha$ approaches faster
to the Haldane limit.
For $\alpha=1$, $O_{\rm S}$ varies gradually through the isotropic
ladder point ($\beta=0$). 
Like $O_{\rm D}$, the string order changes continuously 
keeping finite values from $\beta=-\infty$ to $\beta \sim 1$. 
This strongly suggests that the properties of the Haldane phase 
are kept in almost the whole $\alpha$-$\beta$ plane except 
just on the $\alpha$ axis. 
\par

In fact, similar conclusions have been already reached in the 
case of $\alpha=0$ by the ED and variational 
calculations,\cite{Hida,HidaT} and for $\alpha\ne 0$ by variation 
methods.\cite{Brehmer1,Naka1} 
We demonstrate that the transition from the Haldane phase to 
a spin liquid phase occurs at $\beta=1$ in the case of the SS 
line in Appendix\ref{chap:SS}. 
This discussion can be generalized to all the $\alpha$-$\beta$ plane. 
As we will see later, this ``spin liquid phase" is characterized 
by the delocalized spin polarization, in contrast to the localized 
moment at the edges in the Haldane phase. 
\par

%
\section{Incommensurate Correlation}
In this section we study the incommensurate (IC) correlation 
the model eq.~(\ref{eq:hamil}) presents. 
In \S4.1 and \S4.2 we discuss the IC correlation with respect 
to $S_{ij}$ and $S(q)$, respectively. 
In \S4.3 we construct a comprehensive phase diagram of the 
IC correlation in the $\alpha$-$\beta$ plane. 
\par

\subsection{Disorder in the real space}\label{chap:income1}
In classical spin systems with competing interactions which
favor different wave numbers, \eg $\pi$ and $\pi/2$,
a transition often occurs from an AF ordered state to spiral
ones, as some parameter varies.
Corresponding to this transition, in 1D quantum spin systems 
the oscillatory period of $S_{ij}$ and the wave number 
which maximizes the static structure factor $S(q)$ 
change from a commensurate (C) value to an IC one, although 
LRO's are not realized due to quantum fluctuations. 
Thus, incommensurability is measured in two ways, namely by 
$S_{ij}$ and $S(q)$. 
Generally, the transition point for $S_{ij}$ (disorder point) 
does not coincides with one for $S(q)$ (Lifshitz point).~\cite{Scholl} 
In this subsection, we focus on the former. 
\par

Before discussing our results, we briefly review the classification 
and attribute of C-IC transitions. 
For Ising-spin systems, various properties of C-IC transitions 
were studied; the transitions are classified into two 
kinds.\cite{Stephenson} 
This classification is applicable to the quantum 
systems.\cite{Scholl,Kolezhuk} 
The disorder point of the first kind has the following properties: 
In one side of this point $|S_{ij}|$ has monotonic exponential 
decay, whereas in the other side the exponential decay is 
modified by an oscillatory factor with parameter-dependent 
wavelength, typically as, 
\begin{eqnarray}
\langle S^{z}_{i}S^{z}_{i+r} \rangle 
\sim \cos \left[q(\beta) r \right] e^{-r/\xi(\beta)}.
\label{eq:income}
\end{eqnarray}
In addition, the correlation length $\xi$ behaves in the vicinity 
of this disorder point ($\beta=\beta_{\rm d}$) as, 
\begin{equation}
\left|\frac{\partial\xi}{\partial\beta}\right|_{\rm C}=\infty,\qquad
\left|\frac{\partial\xi}{\partial\beta}\right|_{\rm IC}={\rm finite},
\label{eq:1st}
\end{equation}
where we assume $\beta$ to be the parameter and C (IC) denotes
the limit of $\beta\to\beta_{\rm d}$ in the C (IC) side. 
On the other hand, the disorder point of the second kind has 
the properties that $S_{ij}$ decays exponentially with 
parameter-independent oscillations in both sides of the 
point. 
Besides, $\xi$ satisfies the following conditions:
\begin{equation} 
\xi=0\quad(\beta=\beta_{\rm d}),\qquad 
\left|\frac{\partial\xi}{\partial\beta}\right|_\pm=\infty, 
\label{eq:2nd}
\end{equation}
where the double sign indicates the limits from both sides of
the disorder point.
\par

\begin{figure}
\begin{center}
\epsfxsize=8cm \epsfbox{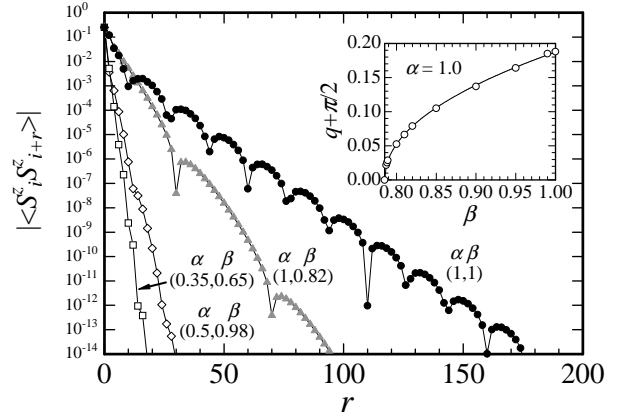}
\end{center}
\caption{
Spin-spin correlation function as a function of $r$ for some 
parameter values in the IC area. 
Only the data of even $r$ are plotted for clarity. 
The inset shows the characteristic $q$ of eq.~(\ref{eq:income}) 
for $\alpha=1$ ($\beta_{\rm d}^{\rm u}=0.775$). 
}
\label{fig:crrincome}
\end{figure}

Now, we consider our model, which has the competing AF couplings, 
the nearest-neighbor and NNN terms. 
As pointed out for Fig.~\ref{fig:crr2}(b), there are two C-IC 
boundaries. 
The lower boundary coincides with the SS line. 
Hereafter, we use $\beta_{\rm d}^{\rm l}$ and $\beta_{\rm d}^{\rm u}$ 
for the lower and upper boundary, respectively. 
As for the C areas, $S_{ij}$ has the 
$\uparrow\uparrow\downarrow\downarrow$ correlation (period 4) for 
$\beta<\beta_{\rm d}^{\rm u}$, and the AF (staggered) correlation 
(period 2) for $\beta > \beta_{\rm d}^{\rm l}$. 
In Fig.~\ref{fig:crrincome}, $|S_{ij}|$ is shown for some typical 
IC cases ($\beta_{\rm d}^{\rm u}<\beta<\beta_{\rm d}^{\rm l}$). 
In contrast to a smooth exponential decay for the C cases 
in Fig.~\ref{fig:fit1}, for $\beta>\beta_{\rm d}^{\rm u}$, 
$|S_{ij}|$ has an oscillation with the long wave length which 
indicates the discrepancy from the period 4 (\ie $q=\pi/2$). 
With increasing $\beta$, this oscillatory period becomes shorter, 
as seen for $\alpha=1$ in Fig.~\ref{fig:crrincome}. 
The inset shows the characteristic wave number $q$ in 
eq.~(\ref{eq:income}), obtained through the following fit:
\begin{eqnarray}
\langle S_{i}^{z}S_{i+r}^{z} \rangle e^{\xi/r} \propto
\cos\left[(q+\pi/2)(r+0.44)\right]. 
\nonumber
\end{eqnarray}
The exponent $\theta$ of 
$q+\pi/2\propto (\beta-\beta_{\rm d}^{\rm u})^{\theta}$ is 
estimated as $\theta=0.39$ for $\alpha=1$, for example. 
On the other hand, the condition for $\xi$ eq.(\ref{eq:1st}) is 
also satisfied, \eg as seen obviously for 
$(\alpha,\beta)=(1,0.775)$ in Fig.~\ref{fig:crr2}(b).
Thus, we have found the upper disorder line belongs to the first 
kind. 
Through the same procedure for $\beta_{\rm d}^{\rm l}$, we have 
confirmed that the lower boundary also belongs to the disorder 
point of the first kind.
By the way, at the isolated dimer point the upper and lower 
boundaries merge into one, and the intermediate IC region 
vanishes [see Figs.~\ref{fig:crr2}(b) and \ref{fig:abplane}]. 
Thus, in both sides of $\beta=0$ the oscillatory periods of 
$S_{ij}$ are constant (2 and 4), and eq.~(\ref{eq:2nd}) is 
satisfied as seen in Fig.~\ref{fig:crr2}(b). 
Accordingly, the isolated dimer point belongs to the second 
kind. 
In this sense, the C-IC transition of the second kind is 
regarded as a complex of two transitions of the first kind. 
\par

Finally, we summarize our results for three typical cases in 
Table~\ref{table:2} with those of the two related $S=1$ spin 
models for comparison. 
\par

\begin{fulltable}
\caption{
Comparison of the C-IC transition among the three models. 
The left column is the $S=1$ bilinear-biquadratic Heisenberg 
chain: 
$
{\cal H} =J\left[\sum_{i}{\mib S}_{i} \cdot {\mib S}_{i+1}
         + \gamma\sum_{i}({\mib S}_{i} \cdot {\mib S}_{i+1})^{2}\right], 
$
where $J>0$ and $\gamma>0$~\cite{Bursill2,Scholl}. 
For $\gamma=1/3$ the model is so-called AKLT Hamiltonian~\cite{AKLT}
and is written in the form of a projection operator. 
The middle column is the $S=1$ frustrated Heisenberg chain: 
$
{\cal H} =J\left[\sum_{i}{\mib S}_{i} \cdot {\mib S}_{i+1}
         +\alpha\sum_{i}{\mib S}_{i} \cdot {\mib S}_{i+1}\right], 
$
where $J>0$ and $\alpha>0$~\cite{Kolezhuk}.
For the present model we tabulate three typical cases.
In the section of $S_{ij}$ the values of the lower and upper 
disorder points are entered with their classes. 
In the section of $S(q)$ the values of the lower and upper 
Lifshitz points are entered with their pitch angles [$q_{\rm max}$] 
in the C sides. 
In the `Other transition' section we list transitions in the
IC regime, if any.
}
\label{table:2}
\begin{fulltabular}{@{\hspace{\tabcolsep}\extracolsep{\fill}}lccccc} \hline
         & $S=1$ bilinear-biquadratic & $S=1$ frustrated &
           \multicolumn{3}{c}{The present model} \\ \cline{4-6}
         & Heisenberg model & Heisenberg model &
           $\alpha$ axis ($\beta=1$)& $\beta$ axis ($\alpha=0$) 
         & $\alpha=0.35$  \\ \hline
$S_{ij}$ \quad\ lower & $\gamma=1/3$ (AKLT) 
& $\alpha=0.284(1)$ & $\alpha=0.5$ (MG) &
           $\beta=0$           & $\beta=0.7$    \\
         &       (first kind)  &   (first kind) &  (first kind) &
                (second kind)  &   (first kind) \\
\qquad\quad upper & $\gamma=1$  & $\alpha=\infty$ ? & $\alpha=\infty$ &
           $\beta=0$           & $\beta=0.445(5)$   \\
         &                     &                &               &
                               &   (first kind) \\ \hline
$S(q)$\quad lower & $\gamma=0.43806(4)$ [$\pi$] & $\alpha=0.373(3)$ [$\pi$]
         & $\alpha=0.52063(6)$ [$\pi$] &
           $\beta=-0.3723(2)$ [$\pi$]       & $\beta=0.5131(1)$ [$\pi$]  \\
\qquad\quad upper & $\gamma=1$ [$2\pi/3$] & $\alpha=\infty$ ? [$\pi/2$] &
           $\alpha=\infty$ [$\pi/2$] &
           $\beta=\infty$ ? [$\pi/2$] & $\beta=\infty$ ? [$\pi/2$] \\
         \hline
$\Delta$(max) & $\gamma=0.406$ & $\alpha=0.40(1)$ & $\alpha\sim 0.6$ &
           $\beta=0$ & $\beta=0.525(5)$ \\
$O_{\rm S}$(max) & $\gamma=1/3$ & $\alpha=0.4397(1)$ &
           $\alpha=0.748(2)$ & 
           $\beta=-4.47(1)$ & $\beta=-2.93(1)$ \\ \hline
Other    & No                  & $\alpha=0.7444$ & No &
           No                  & No \\
Transition &                   & (AKLT--NNN AKLT) && \\ \hline
\end{fulltabular}
\end{fulltable}
%
\subsection{Incommensurate correlation in $S(q)$}\label{chap:income2}
Another measure of incommensurability is the wave number
$q=q_{\rm max}$ which maximizes the static structure factor: 
\begin{eqnarray}
S(q)=\frac{1}{L}
\sum_{i,j}\langle S^{z}_{i}S^{z}_{j}\rangle e^{iq(r_{i}-r_{j})} . 
\label{eq:Sq} 
\end{eqnarray}
We call $q_{\rm max}$ as the pitch angle by analogy with the spiral 
spin order. 
The C-IC boundary (\ie Lifshitz point) is defined as the point at 
which $q_{\rm max}$ deviates from the C wave numbers, $\pi$ or $\pi/2$. 
\par

We have estimated $S(q)$ by DMRG under OBC in the following way: 
Fixing the system size at $L_{0}=280$, we calculate every
$S_{ij}= \langle S^{z}_{i}S^{z}_{j}\rangle$ in the central
part of the system. 
To eliminate the edge effect we do not use 40 sites from both
ends, so that $S_{ij}$'s of $r_{ij}=r_{i}-r_{j}\le 198$ are 
obtained. 
Additionally, thanks to the dimerization, $S_{ij}$'s of negative 
$r_{ij}$ can be estimated only by shifting the starting site 
from $i$ to $i+1$. 
Therefore, we substantially obtain $S_{ij}$ for the system of 
$L=2\times198$. 
Since $S_{ij}$ decays exponentially in the region concerned, 
the long-distance contribution is negligible in the 
Fourier transformation. 
The values of $S(q)$ thus obtained coincide with the ED results 
with PBC up to $L=26$ within 7-8 digits. 
\par

\begin{figure}
\begin{center}
\epsfxsize=8cm \epsfbox{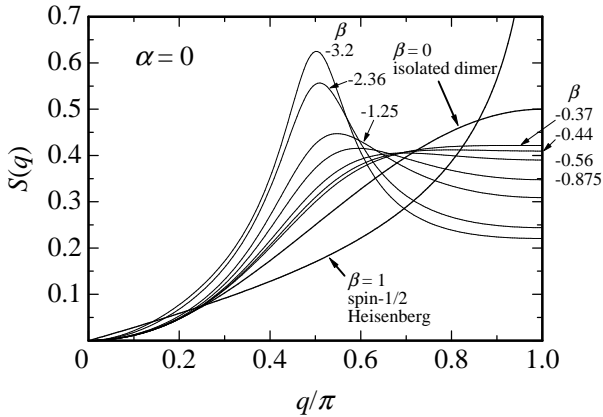}
\end{center}
\caption{
Static structure factor for various values of $\beta$ and 
$\alpha=0$. 
The system used corresponds to the periodic chain of $L=396$ 
(see text). 
Note that the line only connects the discrete data points without
any interpolation. 
}
\label{fig:Sq}
\end{figure}

Figure~\ref{fig:Sq} shows $S(q)$ for various values of $\beta$, 
and $\alpha=0$. 
In the $S=1/2$ Heisenberg model ($\beta=1$), $S(q)$ is 
logarithmically divergent at $q=\pi$ due to the quasi-long-range
AF correlation~\cite{Heisen}. 
At the isolated dimer point ($\beta=0$), $S(q)=[1-\cos(q)]/4$. 
For $1\ge\beta\ge\beta_{\rm L}=-0.3723(2)$, $q_{\rm max}$ remains 
at $\pi$, and starts to shift from $\pi$ at $\beta_{\rm L}$ 
(Lifshitz point). 
In the Haldane limit ($\beta\to-\infty$), $S(q)$ becomes a 
Lorentzian\cite{NomuraS1} sharply peaking at $q_{\rm max}=\pi/2$. 
From the present calculations, however, it is difficult to 
determine definitely the upper C-IC boundary. 
\par
\begin{figure}
\begin{center}
\epsfxsize=8cm \epsfbox{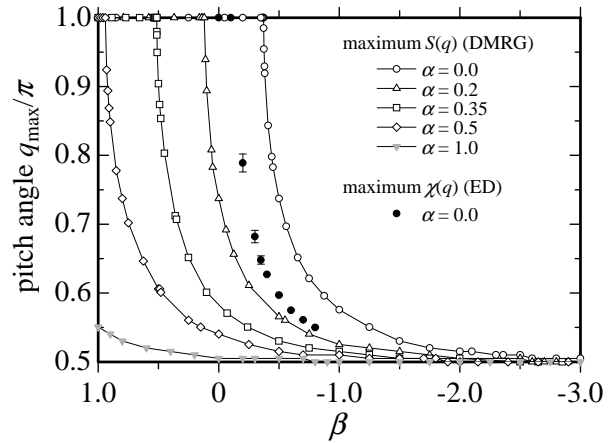}
\end{center}
\caption{
Pitch angle $q_{\rm max}/\pi$ of $S(q)$ as a function of 
$\beta$ for several values of $\alpha$. 
Since the data are of $L=396$, only discrete values of 
$q_{\rm max}$ are allowed. 
Filled circles shows $q_{\rm max}/\pi$ of the static 
susceptibility $\chi(q)$ for $\alpha=0$. 
}
\label{fig:qmax}
\end{figure}

In Fig.~\ref{fig:qmax} we plot $q_{\rm max}$ for some values of 
$\alpha$ versus $\beta$. 
The pitch angle $q_{\rm max}$ suddenly drops from the higher C 
value $\pi$, but slowly approaches the lower C value $\pi/2$, as 
$\beta$ decreases. 
This aspect is common to the $S=1$ models.\cite{Bursill2,Kolezhuk} 
Since for $\alpha>0.52063$\cite{Bursill1} $q_{\rm max}$ is IC even 
for $\beta=1$, $q_{\rm max}$ is always IC for $\alpha=1$. 
In Fig.~\ref{fig:qmax} also shown is $q_{\rm max}$ of the static 
susceptibility $\chi(q)$ for $\alpha=0$, obtained by ED through,
$$
 \chi(q)=\int_0^{\infty}d\omega \frac{S(q,\omega)}{\omega}. 
$$
This quantity is sometimes used to measure the incommensurability,
but the pitch angle of $\chi(q)$ does not necessarily coincide 
with that of $S(q)$. 
\par
In this connection, the tendency of IC peak in $S(q)$ has 
been observed in the inelastic neutron scattering spectrum 
for doped CuNb$_2$O$_6$,\cite{KodamaINS} as discussed in 
Appendix~D.
\par

\subsection{Phase diagram of incommensurate correlation}
\begin{figure}
\begin{center}
\epsfxsize=8cm \epsfbox{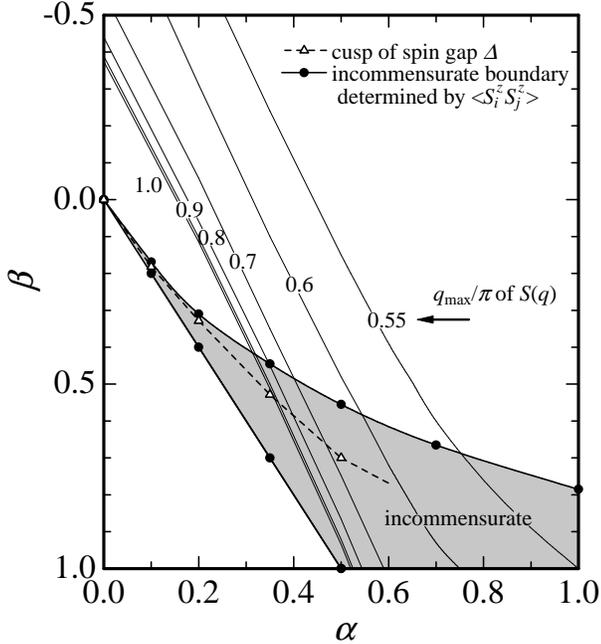}
\end{center}
\caption{
Comprehensive phase diagram of IC correlation in eq.~(\ref{eq:hamil}). 
Shaded area is the IC regime in $S_{ij}$.
Thin solid lines are the contours of pitch angle $q_{\rm max}$ which
maximizes $S(q)$; the Lifshitz line corresponds to the line of 
$q_{\rm max}/\pi=1.0$.
The IC area in $S(q)$ extends further in the directions of 
$\alpha\to\infty$ and $\beta\to -\infty$.
Dashed line with open triangle shows the cusp of the spin gap, 
which distinguishes the wave numbers of the lowest-energy excitation
between $q=\pi$ and $\pi/2$ (see \S3.1).
}
\label{fig:abplane}
\end{figure}

So far, we have studied the IC correlations in terms of three 
quantities: The wave number of the lowest triplet excitation 
$E_{\rm t}(q)$ in \S\ref{chap:sgap}, the spin-spin correlation 
$S_{ij}$ in \S\ref{chap:income1} and the static structure factor 
$S(q)$ in \S\ref{chap:income2}. 
For each quantity we have determined C-IC boundaries in the 
$\alpha$-$\beta$ plane, which are depicted 
in Fig.~\ref{fig:abplane}. 
\par

The dashed line with triangle represents the boundary between 
$q=\pi$ (0) and $\pi/2$ for the lowest $E_{\rm t}(q)$. 
On this line the spin gap shows the cusp as seen in Fig.~\ref{fig:gapb}. 
This jump from $\pi$ (0) to $\pi/2$ with decreasing $\beta$, 
\ie cusp, is obvious for $\alpha \lsim 0.5$, but it disappears 
for $\alpha\gsim 0.5$ due to the IC excitations (see 
Fig.~\ref{fig:cusp} in Appendix\ref{chap:A-B}). 
This line is included in the IC region of $S_{ij}$, and discords 
from each boundary mentioned below. 
\par

The shaded area is the IC region where $|S_{ij}|$ decays 
exponentially with a quasi-periodic oscillation as in 
Fig~\ref{fig:crrincome}. 
This region is bounded by the disorder lines indicated by the 
solid lines with filled circle. 
The C region under the lower (over the upper) 
disorder line has the AF ($\uparrow\uparrow\downarrow\downarrow$) 
correlation. 
On the lower boundary (the SS line), the Hamiltonian can be 
written in the form of projection operators.\cite{AKLT} 
The fact that $\cal H$ is represented by projection 
operators at the disorder point is common to some other models. 
In the spin-orbit model the disorder point is the SU(4) 
point,\cite{Pati} and in the $S=1$ model the AKLT point,\cite{Scholl}
as shown in Table~\ref{table:2}.
\par

The thin solid lines show the contours of the pitch angle 
$q_{\rm max}/\pi$, which maximizes $S(q)$. 
In contrast to other cases,\cite{Scholl,Kolezhuk} the IC region 
of $S(q)$ extends widely in the C region of $S_{ij}$. 
The pitch angle decreases abruptly near the Lifshitz line 
($q_{\rm max}=\pi$). 
In such a region $S(q)$ shows merely a broad hump instead of
a sharp peak (Fig.~\ref{fig:Sq}), hence $q_{\rm max}$ is hardly 
regarded as a characteristic wave number there. 
Indeed, we have not observed a singular behavior in any quantity
at the Lifshitz point; system-size dependence of many quantities 
in this IC region shows the period of 4, in accordance with 
the C period of $S_{ij}$. 
Thus, the intrinsic IC correlation is represented in terms of 
$S_{ij}$.
\par

In Fig.~\ref{fig:abplane} all the lines go down rightward. 
As repeatedly mentioned, this indicates that the properties 
are preserved along this direction due to the similar role 
of $-\beta$ and $\alpha$. 
\par

Lastly, we touch on the continuity of the Haldane phase. 
The IC regime lies between the Haldane limit ($\beta=-\infty$) 
and $\beta=1$ irrespective of the value of $\alpha$.
Indeed, the correlation length and the spin gap are singular 
there. 
However, various quantities including the dimer and string 
order parameters (\S\ref{chap:order}), which characterize 
the Haldane phase, have no anomaly. 
Actually, we will confirm that the Haldane phase extends to 
the limit of $\beta\to 1$ in \S5 and Appendix C. 
Hence, the IC correlation does not affect the essence of 
the Haldane phase and also of the spin liquid phase. 
\par
%
\section{Spin-Peierls Regime}
In this section, we concentrate on the Spin-Peierls regime. 
First, the low-lying excitation is reconsidered in \S5.1. 
Next, we study the edge effects in the light of the delocalized 
spin polarization, and consider the origin of the impurity effect 
of CuGeO$_3$ in \S5.2. 
The stability of the edges is considered in reference to the 
lattice distortion in spin-Peierls compounds in \S5.3.
The NMR spectra is discussed in \S5.4.
\par

\subsection{Two-spinon bound state}
One primitive modeling of the substitution by non-magnetic 
elements is to consider finite-size systems, assuming that 
the impurities simply cut the 1D chain into finite segments. 
For finite systems, the boundary conditions are essentially 
important, in particular for the dimerized cases ($\beta< 1$). 
Then, there are two kinds of exchange coupling at the chain 
ends: $J$ (stronger one) or $\beta J$ (weaker one). 
For simplicity, we call the former (latter) case as the strong 
(weak) edge. 
In even-site systems both of the edge couplings are $J$ or 
$\beta J$. 
\par

The local spin polarization $\langle S^{z}_{i}\rangle$ is an 
important quantity to study the local properties of magnetic 
excitation and the tendency toward magnetic LRO's.\cite{Sznote} 
Figure~\ref{fig:Sz} shows $\langle S^{z}_{i}\rangle$ of the 
lowest-energy state in the $S^{z}=-1$ subspace for $\alpha=0$ 
(I) and $\alpha=0.35$ (II). 
For both values of $\alpha$ the cases of the strong and weak edge 
are shown in (a) and (b), respectively. 
As described in Appendix~C.2, the strong edge has some analogous 
properties to PBC or to the bulk. 
Since the main subject of this subsection is the low-lying 
excitation in the bulk, we restrict our discussions to the 
qualitative feature of the strong edge, here. 
We will concentrate on the edge properties and impurity effect 
in \S5.2-\S5.4. 
\par

In Figs.~\ref{fig:Sz} I(a) and II(a) there are two humps of 
staggered moment, and the amplitude becomes maximum for $\beta=1$. 
These humps are considered as the amplitude of excited free 
spinons ($S=1/2$ entity).\cite{SorenAAP,note4spinon} 
For the $S=1/2$ Heisenberg model [($\alpha$,$\beta$) =(0,1)], 
it is known that the excitation can be exactly described by the 
scattering of even number spinons~\cite{Faddeev,Bougourzi}. 
For the MG point [($\alpha$,$\beta$)=(0.5,1)], a similar free 
spinon picture was successfully introduced from a variational 
viewpoint,\cite{SS} and recently it was numerically shown that 
the MG gap (two-spinon excitation) is accurately twice the bottom 
of the single spinon blanch for odd-site systems~\cite{SorenAAP}. 
Thus, the free-spinon picture seems applicable to all the values 
of $\alpha$ for $\beta =1$ at least in the low-lying excitation. 
\begin{fullfigure}
\figureheight{9cm}
\begin{center}
\end{center}
\caption{
Local spin polarization of the lowest-energy state in the
subspace of $S^{z}=-1$ for open systems with $L$=200 as a
function of $i$ for $\alpha$=0 (I) and $\alpha$=0.35 (II). 
At both chain ends the strong coupling $J$ is assumed in (a),
while the weak coupling $\beta J$ in (b).
For $\beta\sim 1$ in every panel, $\langle S^z_{i}\rangle$ 
oscillates alternately with considerable amplitude.
}
\label{fig:Sz}
\end{fullfigure}
%
When the dimerization is switched on ($\beta <1$), the amplitude 
of the oscillating part is abruptly reduced, and almost disappears 
for $\beta=0.2$ leaving a broad peak as seen in Figs.~\ref{fig:Sz} 
I(a) and II(a). 
The mechanism of this process has been pursued by many 
studies.\cite{Tsvelik,Uhrig,Poilblanc,Brehmer2,SorenAAP} 
When $\beta$ decreases from 1, two spinons are bound to make a 
magnon ($S=1$ entity) described by a single parameter; an isolated 
blanch thereby appears below the continuum. 
Furthermore, multi-magnon branches have been found above
it.\cite{Bouzerar,YS,SorenAAP} 
\par

\begin{figure}
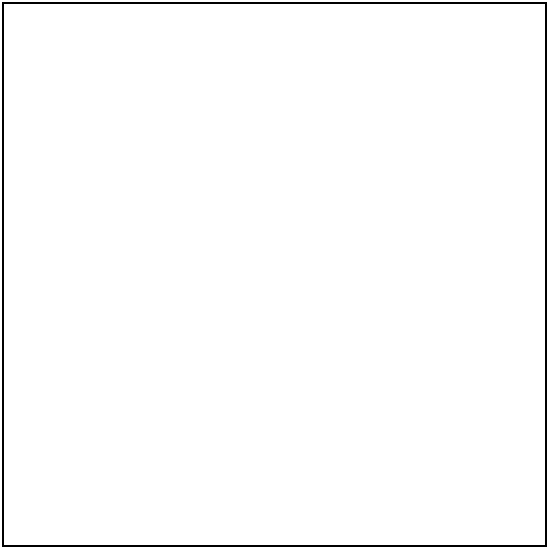

\figureheight{7cm}
\begin{center}
\end{center}
\caption{
Schematic figures for low-energy excitations along the SS line 
($\beta= 2\alpha$).
I(a),(b): The degenerate ground states for the MG point.
II: The lowest-excited state at the MG point, in which two spinons
on different subchains (legs) form domain walls.
III(a),(b): Excited states on the SS line ($\beta<1$), where two 
spinons are located apart and closely, respectively. 
See text for details.
}
\label{fig:Szpic}
\end{figure}

To grasp intuitively the process from spinon to magnon, it is
useful to consider the excitation along the SS line. 
At the MG point the ground state is twofold degenerate, 
$\psi_1$ and $\psi_2$ in eq.~(\ref{eq:psi1}).
Here, we call the singlet array of $\psi_1$ ($\psi_2$) as
the R (L) array [Figs.~\ref{fig:Szpic} I(a) and I(b)].
In the lowest-energy excitation a singlet pair breaks into
two spins (spinons), which sit on the mutually different subchains 
(legs) to avoid breaking another singlet pair, as shown in 
Fig.~\ref{fig:Szpic} II.
The resultant spinons become two domain walls, between which
the phase of singlet array, say L, is opposite to the original
one, R.
In this case the L array has an equivalent energy to the R array, 
so that the spinons move freely regardless of the distance 
between them.
Thus, these scattering states make a low-energy continuum.
On the other hand, when $\beta$ is reduced from 1 keeping 
$\beta=2\alpha$, the degeneracy of the ground state is lifted.
One of the states, say $\psi_1$ in which singlets are positioned
on the strong bonds ($J$) is still the exact ground state for 
$\beta < 1$, while $\psi_2$ is no longer an eigenstate and even 
has high energy over the singlet-triplet gap, as explained 
in Appendix C. 
Consequently, the L array is energetically unfavorable and works
as a linear confinement potential; the two spinons approach each 
other to reduce the region of L. 
Namely, the configuration of Fig.~\ref{fig:Szpic} III(b) is more 
likely than that of Fig.~\ref{fig:Szpic} III(a). 
The resultant bound state causes an isolated branch below the 
continuum. 
Although we have focused on the SS line, this explanation is 
qualitatively valid for all the other paths from the $\alpha$ 
axis. 
This change of the excitation mechanism corresponds to the 
transition from the spin liquid phase to the Haldane phase,\cite{Brehmer2} 
and is analogous to the so-called $\Delta$ chain.\cite{Delta}
\par
%
Actually, one can confirm the appearance of the isolated magnon
branch in the dynamical structure factor, 
\begin{equation}
S(q,\omega)=\sum_n |\langle\Psi_n|S_q^z|\Psi_0\rangle|^2
\delta\left(\omega-(E_n-E_0)\right),
\end{equation}
where $\hbar=1$, $\Psi_n$ ($E_n$) is the $n$-th eigenfunction
(eigenvalue) of the system, and the ground state is indicated by $n=0$.
$n$ runs over all the eigenstates.
$S(q,\omega)$ is proportional to the intensity of inelastic 
neutron scattering spectra. 
In Figs.~\ref{fig:sqwcgo}(a)-(c) we show $S(q,\omega)$ for 
$\tilde\alpha=0.35$, which is the plausible value for 
CuGeO$_3$.\cite{notealpha}
As discussed in the previous paper,\cite{YS} the lowest mode
for $\delta=0$ ($\beta=1$) [Fig.~\ref{fig:sqwcgo}(a)] 
is assigned to the 
continuum by checking the system-size dependence of the residue 
and position of pole in $S(q,\omega)$.
As soon as the dimerization is introduced, weak intensity appears
below the continuum, whose position is almost independent of $L$,
as seen in Fig.~\ref{fig:sqwcgo}(b).
This branch remarkably grows when $\delta$ increases to 0.012 
($\beta=0.98$) [Fig.~\ref{fig:sqwcgo}(c)], which is an expected 
value for CuGeO$_3$ within eq.~(\ref{eq:hamil}).\cite{Bouzerar,Poilblanc}
This abrupt increase of the intensity is obvious in the inset
of Fig.~\ref{fig:sqwcgo}(b), where the ratio of intensity 
(residue) for the lowest branch at $q=\pi/2$ is plotted.
Since the intensity belonging to the continuum decreases with
$1/L$ as $L\to\infty$, this isolated magnon branch makes a 
dominant contribution.
\par
\begin{fullfigure}
\figureheight{7cm}
\begin{center}
\end{center}
\caption{
Dynamical structure factor $S(q,\omega)$ for three values
of $\delta$ are shown for the fixed value of $\tilde\alpha=0.35$.
Data obtained by the ED with a recursion method for $L=14$-$26$
are simultaneously plotted.
The intensity is proportional to the area of circle.
The magnon branch is indicated by arrow in (b) and (c).
The inset in (b) shows the intensity ratio (residue) of the
lowest branch at $q=\pi/2$ versus $\delta$ for $L=20$ and $24$.
System-size dependence is small in this wave number. 
}
\label{fig:sqwcgo}
\end{fullfigure}
%
Recent neutron experiments\cite{Fujita} for CuGeO$_3$ vividly 
showed the appearance of the magnon branch. 
In the measurement at $T=12$K (immediately below 
$T_{\rm SP}\sim 14$K) the spectrum has a sharp peak below the 
continuum as previously observed,\cite{Ain} 
whereas at $T=16$K this sharp peak suddenly vanishes and only 
a broad continuum is observed. 
Thus, the change in the excitation mechanism between $\delta=0$ 
and $\delta>0$ has been confirmed by experiment.
\par

Lastly, we touch on the thermodynamics. 
According to the abrupt change of the level structure near 
$\delta=0$, thermodynamic quantities drastically changes, too. 
We 
confirmed that by calculating 
the specific heat $C(T)$ along the SS line. 
As soon as $\tilde\alpha$ decreases from 0.5, a sharp peak 
appears. 
This peak is attributed to the dense low-lying triplet levels 
caused by the dimerization. 
The situation is the same with the $\Delta$ chain.\cite{Delta} 
\par
%
%
\subsection{Delocalized spin polarization}
In this subsection, we study $\langle S^{z}_{i}\rangle$ in the 
lowest magnetic state ($S\ne 0$) to clarify the impurity effect. 
We assume that non-magnetic impurities cut the chain at random 
into odd- and even-site segments. 
This assumption is not so crude, since the NNN exchange hardly 
change the situation, as will be discussed later. 
\par
\par

\begin{figure}
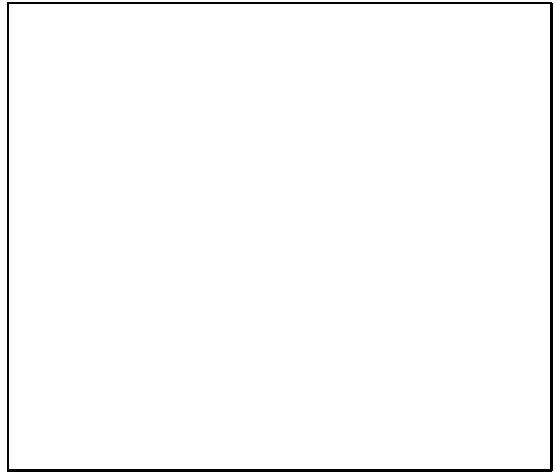

\figureheight{6cm}
\begin{center}
\end{center}
\caption{
Local spin polarization in the ground state ($S^z=1/2$) for 
various cases. 
(a) Segment with odd sites for some values of $\beta$. 
A weak edge ($\beta J$) is assumed in the left chain end.
(b) Segment with $L=21$ in the IC regime without the
coupling alternation. 
(c) System with $L=24$ in which the 12th site is replaced by 
a site impurity. 
See text for details. 
In the margin of (a) and (c), the respective connectivity is 
schematically shown. 
As for (b), the connectivity is the same with (a). 
}
\label{fig:Szodd}
\end{figure}

First, we look at the odd-site systems. 
Although for even-site segments the ground state is always singlet 
and $\langle S^{z}_{i}\rangle=0$ for each $i$, yet for odd-site 
segments the ground state is $S=1/2$, therefore 
$\langle S^{z}_{i}\rangle\ne 0$. 
In Fig.~\ref{fig:Szodd}(a) we show $\langle S^{z}_{i}\rangle$ 
for $L=21$ and $\alpha=0.35$ in the $S^z=1/2$ sector. 
For the non-dimerizing case ($\beta=1$), the spin polarization 
extends to the whole system, corresponding to a spinon discussed 
in \S5.1. 
This delocalized spin polarization is still seen in the IC 
regime ($\alpha>0.5$) as far as $\beta=1$, as shown in
Fig.~\ref{fig:Szodd}(b). 
Therefore the delocalized spin polarization is characteristic 
of the spin-liquid phase, and has nothing to do with the 
existence of a gap. 
For $\beta<1$ one of the chain ends is the strong edge 
[$i=21$-$22$ in Fig.~\ref{fig:Szodd}(a)] and the other the 
weak edge ($i=1$-$2$). 
With decreasing $\beta$, the magnetic moment becomes localized 
near the weak edge and the amplitude near the strong edge vanishes. 
Here, let us return to Figs.~\ref{fig:Sz} I(b) and II(b), where the 
lowest excited states ($S=1$) in even-site segments with the weak 
edges are shown. 
These states are obviously regarded as the duplication of the ground 
state of odd-site systems ($S=1/2$) connected at the strong edges. 
Hence, the localization of magnetic polarization is a characteristic 
of the weak edge. 
This free $S=1/2$ excitation at the edge is nothing but the 
celebrated aspect of the Haldane chain.\cite{Kennedy,MiyashitaY} 
\par

\begin{figure}
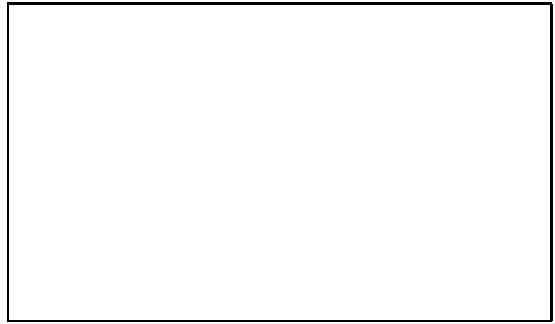

\figureheight{4cm}
\begin{center}
\end{center}
\caption{
Amplitude (maximum value) of $\langle S^z_i\rangle^{\rm s}$ for 
the strong edge, and the range of localized moment near the chain
ends $\eta_{\rm edge}$ for the weak edge, as a function of $\beta$. 
The inset shows the system-size dependence of the uniform part 
(value at the mid-chain site) and the staggered part (maximum value) 
of $\langle S^z_i\rangle$ for the Heisenberg model. 
The used states have the lowest energies in the $S^z=-1$ subspace.
}
\label{fig:Szsize}
\end{figure}

In the following, we treat this localized moment more quantitatively. 
Consider a finite system with $L$ spins, as in Figs.~\ref{fig:Sz} I(b) 
and II(b). 
For the non-dimerized case ($\beta=1$), the range of finite 
$\langle S^{z}_{i}\rangle$ extends to the whole chain irrespective 
of the value of $\alpha$, as mentioned. 
When dimerization is introduced, this range shrinks toward the edges. 
Incidentally, in the spin-Peierls regime ($\beta\sim 1$), 
$\langle S^{z}_{i}\rangle$ around the edges does not obey the simple 
exponential decay unlike the Haldane regime, so that the localization 
length is not an appropriate measure of the localization range. 
Therefore, we introduce a somewhat artificial quantity $\eta_{\rm edge}$, 
defined as the maximum integer satisfying 
$|\langle S^{z}_{i}\rangle|>10^{-3}$ for all $i<\eta_{\rm edge}$. 
Here, we choose the value $10^{-3}$ merely for convenience. 
In Fig.~\ref{fig:Szsize} we show $\eta_{\rm edge}$ estimated 
for the systems of $L=200$. 
Note that $\eta_{\rm edge}$ scarcely depends on the system size, 
in contrast to $\langle S^z_i\rangle$ for the strong edge. 
The localization range abruptly becomes narrow with decreasing 
$\beta$ from 1. 
With increasing $\alpha$, $\eta_{\rm edge}$ behaves more abruptly. 
For $\beta\lsim 0.9$, $\langle S^{z}_{i}\rangle$ around the edges 
shows the typical exponential decay characteristic of the 
Haldane phase. 
\par
%
Next, returning to the Figs.~\ref{fig:Sz} I(a) and II(a), 
we give quantitative treatment of the strong edge. 
In this case the local moment can be decomposed into the uniform 
and the staggered parts as,\cite{Laukamp,Augier}
\begin{equation}
\label{eq:szdec}
\langle S^z_i\rangle=\langle S^z_i\rangle^{\rm u}+
(-1)^i\langle S^z_i\rangle^{\rm s}.
\end{equation}
For $\alpha=0$ the uniform part $\langle S^z_i\rangle^{\rm u}$ 
is almost constant especially in the bulk for $\beta=1$; 
$\langle S^z_i\rangle^{\rm u}$ abruptly deforms with decreasing 
$\beta$, and converges to a sinusoidal curve for $\beta\to 0$. 
For instance, see $\beta=0.2$ in Figs.~\ref{fig:Sz} I(a). 
Since the amplitude of the curve is determined by the magnetization 
and the system size, $\langle S^z_i\rangle^{\rm u}\propto 1/L$ 
as shown in the inset of Fig.~\ref{fig:Szsize}. 
On the other hand, the staggered part $\langle S^z_i\rangle^{\rm s}$ 
is drastically suppressed with decreasing $\beta$ and vanishes 
at $\beta=0$, as shown in Fig.~\ref{fig:Szsize}. 
Furthermore, it is important that $\langle S^z_i\rangle^{\rm s}$ 
highly depends on the system size, and vanishes as $L\to\infty$ 
(inset of Fig.~\ref{fig:Szsize}). 
Thus, $\langle S^z_i\rangle\sim\langle S^z_i\rangle^{\rm u}$ 
for $\beta\sim 0$.
The behaviors of $\langle S^{z}_{i}\rangle$ is basically the same 
for $\alpha=0.35$. 
\par

Based on the above results, we consider the impurity effects in 
the spin-Peierls regime, especially with CuGeO$_3$ 
[$\alpha=0.35$, $\beta=0.97$-$0.98$ for $T<T_{\rm SP}$] in mind. 
For pure CuGeO$_3$ a magnetic LRO has never been observed, 
while for a doped sample with 0.12\% of Zn shows AFLRO in the 
low temperature.\cite{Manabe}
\par

[1] Infinite systems ($L=\infty$, pure material) belong to the Haldane 
phase, as discussed in Appendix C.
(a) For the open strong edge, the staggered polarization extends 
to the whole chain, however its amplitude becomes infinitesimally 
small as $L\to\infty$, as shown in Fig.\ref{fig:Szsize}.
Thus a magnetic LRO will hardly arise, even if the interchain 
coupling is somewhat large.
(b) For the open weak edge, there appears staggered polarization 
with considerably large amplitude in the range of about 30 sites 
around the edges, but this range is negligible against $L=\infty$. 
The bulk area is not affected by this moment; the higher dimensional
coupling will hardly induce a magnetic LRO also in this case. 
Thus, regardless of the kind of edge, infinite (pure) systems will 
keep the gap properties of the Haldane phase (see \S6). 
\par

[2] Finite systems (impurity-doped material) may break the 
characteristics of the Haldane phase (Appendix C).
(a) For the open strong edge, the amplitude of 
$\langle S^{z}_{i}\rangle^{\rm s}$ in the whole system increases with 
the system size decreasing, in particular abruptly for the thin 
impurity concentration (inset of Fig.~\ref{fig:Szsize}). 
Thus, making use of the enhanced staggered moment, the interchain 
coupling probably induce AFLRO.
(b) For the open weak edge, with decreasing the system size, 
the bulk part with $\langle S^{z}_{i}\rangle=0$ vanishes, since 
$\eta_{\rm edge}$ is independent of $L$. 
As a result, the localized polarization around the edges extends to 
the whole system. 
AFLRO is likely to appear also in this case. 
\par

To summarize, when the system becomes finite by doped with non-magnetic 
impurities, the properties of the spin liquid are brought about 
to the spin-Peierls regime, which originally belongs to the Haldane 
phase for the pure case, and AFLRO will be induced by the 
interchain magnetic coupling. 
The above results are not qualitatively influenced by the value of 
$\alpha$, so that the same mechanism works for spin-Peierls 
materials in general. 
So far, we have not made allowance for the stability of the edges. 
In fact, the strong edge is always stable against the weak edge 
in the spin-Peierls regime.
We will discuss this point in the next subsection. 
\par

In this connection, recent $\mu$SR measurements for doped 
CuGeO$_3$ have observed the spatial inhomogeneity of the ordered 
moment size.\cite{KojimaMSR} 
However, its local structure has not been elucidated yet. 
Meanwhile, concerning $\langle S^z_i\rangle$, the result of 
the phase Hamiltonian approach\cite{Fuku} is similar to the
weak-edge case, if anything. 
\par

Finally, we mention the effect of the NNN coupling across the impurity. 
Consider a system consisting of 24 sites but the spin on the 12th site 
is replaced by a site impurity like Zn in CuGeO$_3$ 
[Fig.~\ref{fig:Szodd}(c)]. 
Thereby, the system is separated into odd-site (11) and even-site (12) 
segments by the impurity, but the NNN exchange ($\alpha=0.35$) 
connects them.
Since this system has odd spins (23), the ground state is magnetic 
($S=1/2$). 
In Fig.~\ref{fig:Szodd}(c) we shows a typical result among various 
coupling conditions calculated. 
Here, we set the weak edge in the odd-site segment next to the impurity 
(between $i=10$ and $11$) and the strong edges for the even-site segment. 
An important aspect of this result is that the large amplitude of 
$\langle S_{i}^{z}\rangle$ in the odd-site segment scarcely extends 
to the even-site segment beyond the impurity. 
This aspect does not change under other coupling conditions and also 
in the cases of a bond impurity like Si in CuGeO$_3$. 
Summing up, the NNN coupling does not severely affect our simple
treatment of the impurity effect. 
\par

\subsection{Stability of the edges}
An important factor for the impurity effect in spin-Peierls 
compounds is the softness of the lattice. 
In the low temperature the lattice distorts so as to minimize 
the sum of the lattice distortion energy and magnetic one. 
We can allow for this factor to some extent by considering the 
stability among different edge conditions. 
\par

\begin{figure}
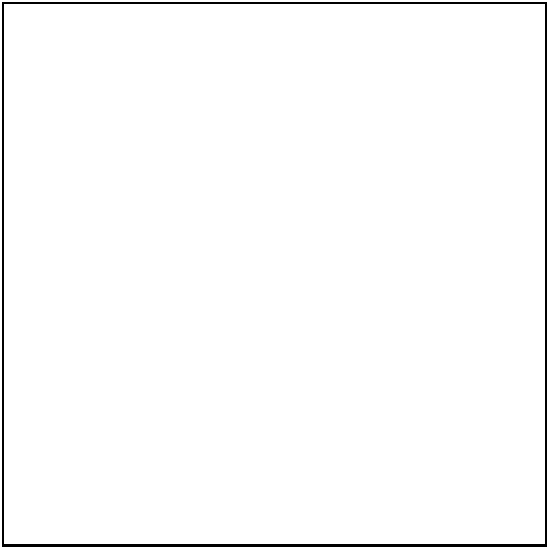

\figureheight{7cm}
\begin{center}
\end{center}
\caption{
(a) Energy difference of the ground state between the two edge 
conditions, which are different between the even and odd 
segments, as shown schematically in (b). 
For details see text. 
In the inset of (a) plotted is the inverse of 
$J\partial\Delta E/\partial\delta$ estimated by small values 
of $\delta$. 
}
\label{fig:SWedgeE}
\end{figure}

First, we consider the energy difference between the ground states 
of the strong and the weak edges in even-site systems. 
In Fig.~\ref{fig:SWedgeE}(a) we plot 
\begin{equation}
\Delta E(L)=\left\{
\begin{array}{ll}
E({\rm SS},L)-E({\rm WW},L), & \mbox{even}\ L \\
E({\rm SS},L)-E({\rm SW},L), & \mbox{odd}\ L 
\end{array}
\right.
\label{eq:SWedgeE}
\end{equation}
versus $\delta$ for $\tilde\alpha=0.35$.
Here, $E({\rm SS},L)$ means the ground-state energy of the system 
with $L$ sites and the strong edges [cf. Fig.~\ref{fig:SWedgeE}(b)]. 
In this range of $\delta$ the energy of the strong edges is 
always lower than that of the weak edges. 
The behaviors of $E({\rm SS},L)$ and $E({\rm WW},L)$ are identical 
with those along the SS line. 
Namely, $E({\rm SS},L)$ is a monotonically decreasing function of 
$\delta$ and proportional to $L$, whereas $E({\rm WW},L)$ highly 
depends on $L$ for $\delta\sim 0$, and has discontinuity at $\delta=0$ 
for $L\to\infty$. 
Therefore, $\Delta E(\infty)$ in the limit of $\delta\to 0$ is 
finite and equivalent to the gap at $\delta=0$. 
For details, refer to Appendix C. 
The finite $\Delta E(\infty)$ for $\delta\to 0$ in this case is recognized 
in the inset, where $\partial\Delta E/\partial\delta$ diverges for 
$L\to\infty$. 
Next, for odd $L$ we also plot $\Delta E(L)$ of eq.~(\ref{eq:SWedgeE}) 
in Fig.~\ref{fig:SWedgeE}(a). 
Here, $E({\rm SW},L)$ is the ground-state energy of regular 
odd chain, while $E({\rm SS},L)$ is that of the strong edges at 
both chain ends [See Fig.~\ref{fig:SWedgeE}(b)]. 
The defect (irregular pattern) of exchange coupling arising 
in the latter is placed at (or near) the center of the chain. 
Since $\Delta E$ is negative even in this case, the strong edge 
lowers the energy, overcoming the energy loss by the defect. 
Incidentally, $E({\rm WW},L)$ is much higher than both energies 
above. 
Furthermore, concerning the first excited (triplet) state, we 
compare the various cases of Figs.~\ref{fig:Sz}(a) and (b) ($L=200$). 
Also in these cases the strong edge has lower energy than the 
weak edge. 
\par

Thus, we conclude that in the spin-Peierls regime both energies 
of the ground state and of the lowest excited state are stabilized 
by the strong edges. 
In other words, if the lattice is sufficiently flexible like 
spin-Peierls compounds, the edge becomes strong. 
We would like to emphasize that this conclusion holds even when 
the elastic energy of the 3D lattice is included. 
Because, as is known, the decrement of magnetic energy due to the 
spin-Peierls distortion $E_{\rm mag}\propto\delta^\nu\ (\nu\le 4/3)$ 
overcomes the increment of elastic energy of the lattice 
$E_{\rm lat}\propto\delta^2$ for $\delta\to 0$.\cite{CrossF,YS}  
\par

The result that the strong edge is favorable is consistent with 
the enhancement of singlet correlation at the edge bond,\cite{Delta,Laukamp} 
although this is not the case for large negative values of 
$\beta$, as will be considered in \S6. 
In the above we have discussed simply with fixed-frame lattices. 
To treat the spin-Peierls transition more accurately, one needs 
to introduce the magneto-elastic coupling. 
However, the results of recent studies\cite{Augier} allowing for 
such a factor are basically the same with ours. 
\par
Based on the above results, we speculate on the first-order 
compositional phase transition in CuGeO$_3$ in the remainder 
of this subsection. 
Recently, Masuda \etal\cite{Masuda} reported that when non-magnetic 
impurity Mg is doped as Cu$_{1-x}$Mg$_x$GeO$_3$, a clear first-order 
transition is observed at $x=x_{\rm c}\sim 0.023$ (later modified 
as 0.027\cite{Nakao}) between the dimerized and uniform AF ordered 
phases. 
The dependence of $T_{\rm N}$ on $x$ has a discontinuity at 
$x_{\rm c}$ and the spin-Peierls order disappears for $x>x_{\rm c}$. 
A recent neutron scattering study\cite{Nakao} has supported this 
first-order transition. 
According to it, the amplitude of magnetic moment $\mu_{\rm eff}$ 
as well as that of the spin-Peierls distortion have jumps at 
$x_{\rm c}$. 
\par

For $T<T_{\rm SP}$ the lattice of CuGeO$_3$ distorts in every 
direction of the three crystal axes\cite{Hirota} (The c axis is 
the chain direction). 
The array of bond alternation in a chain is ``out of phase" against 
those in the adjacent chains in both a- and b-axis directions, 
namely -L-R-L-R- in the notation of Fig.~\ref{fig:Szpic}\ I. 
When a non-magnetic impurity cuts the chain, the two edge bonds 
next to the impurity have to be strong. 
Thus, the array in one side of the impurity becomes ``in phase" 
locally, \eg -L-L-L-. 
This ``in-phase" array raises the elastic energy, so that 
this reversed phase should be relaxed to the original ``out-of-phase" 
array in some distance from the impurity.\cite{Khomskii} 
Actually, it is observed that the width of spin-Peierls superlattice 
peaks in the neutron scattering spectra shows the resolution 
limit,\cite{Martin} namely the intrachain phase of dimerization 
does not change globally across the impurity. 
\par

When the concentration of impurity is thin, the contribution of 
the above local defects to the total energy is smaller than the 
energy gain by the global spin-Peierls distortion. 
As $x$ increases, local defects appear here and there, which 
damage the latter energy gain. 
Eventually, for $x>x_{\rm c}$ the contribution of the defects 
overcomes the energy reduction by the global dimerization.
Thus, the compositional transition of the first order may be 
understood by the competition between the 3D global spin-Peierls 
distortion and the 1D local defects.\cite{noteHirota} 
The discontinuous increase of $\mu_{\rm eff}$ at $x=x_{\rm c}$ 
is understood naturally; if the dimerization vanishes by the 
above first-order transition, $\langle S^z_i\rangle$ is enhanced 
discontinuously. 
For instance, imagine that $\beta$ jumps from 0.99 to 1.0 
in Fig.~\ref{fig:Sz}\ II(a). 
\par
%
\subsection{NMR spectra}
In this subsection we discuss the impurity effect on the NMR 
spectrum $g(H)$. 
Since non-magnetic states ($S=0$) do not contribute to the local 
susceptibility, 
\begin{equation}
\label{eq:chii}
\chi_i=\frac{1}{T}\sum_j\langle S^z_iS^z_j\rangle_T,
\end{equation}
$\chi_i$ is proportional to $\langle S^z_i\rangle$ with respect 
to the lowest magnetic state in the $T\to 0$ limit. 
In eq.~(\ref{eq:chii}) $\langle\cdots\rangle_T$ indicates the 
thermal average. 
Then, by replacing $\chi_i(T)$ by $\langle S^z_i\rangle$ the NMR 
spectra is expressed as 
\begin{equation}
\label{eq:NMR}
g(\tilde H)=\sum_i f\left[\tilde H-\langle S^z_i\rangle\right],
\end{equation}
where $f(H)$ is the NMR line shape with natural line width
at each site, and we assume $f(H)=\exp(-H^2/w^2)$ in this paper.
And $\tilde H=H/(Ag\mu_{\rm B})$ with $A$ being the hyperfine
coupling of the material.
Like eq.~(\ref{eq:szdec}), local susceptibility is sometimes decomposed 
into the uniform and staggered parts:
\begin{equation}
\label{eq:chidec}
\chi_i=\chi_i^{\rm u}+(-1)^i\chi_i^{\rm s}.
\end{equation}
When $\chi_i^{\rm u}$ is independent of $i$, it is nothing but
the bulk susceptibility $\chi=\chi_i^{\rm u}$ and only shifts $g(H)$.
Actually, $\chi_i^{\rm u}$ often depends weakly on $i$ as in
Fig.~\ref{fig:Sz}(a).
On the other hand, $\chi_i^{\rm s}$ determines the line shape.
When $\chi_i^{\rm s}$ has a different oscillatory period from $\pi$, 
\eg for the IC or $\uparrow\uparrow\downarrow\downarrow$ structure, 
the decomposition of eq.~(\ref{eq:chidec}) is unavailable.
\par

First, we look at the Heisenberg model ($\alpha=0$,\ $\beta=1$).
Eggert and Affleck have obtained $\chi_i^{\rm s}(T)$ for OBC by the 
conformal field theory:\cite{EggertA} 
\begin{equation}
\label{eq:chiHeis}
\chi_i^{\rm s}(T)=\frac{2a}{\pi}\frac{i}{\sqrt{(J/2T)\sinh{(4Ti/J)}}},
\end{equation}
with a constant $a\sim 0.58$.
The chain end is at $i=0$.
In relatively high temperatures $\chi_i$ has finite amplitude 
only near the chain end, whereas with decreasing temperature 
the oscillation extends to the inner part of the chain and its 
amplitude increases. 
This aspect is consistent with the delocalization of 
$\langle S^z_i\rangle$ for $T\to 0$ and $\beta=1$. 
Those changes cause the broadenings of NMR line shape and of 
background intensity with sharp edges, respectively. 
Such behaviors were actually observed by an NMR experiment for 
a good Heisenberg compound Sr$_2$CuO$_3$.\cite{Takigawa}
\par

\begin{figure}
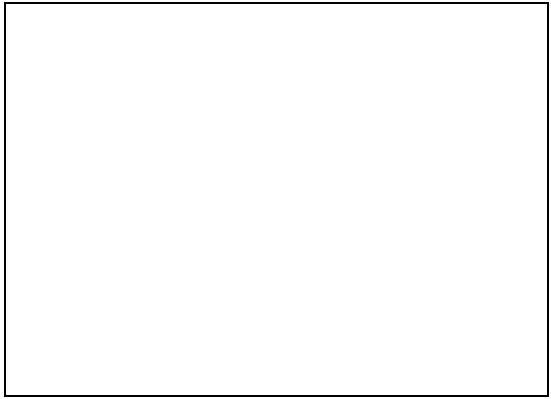

\figureheight{5cm}
\begin{center}
\end{center}
\caption{
System-size dependence of NMR intensity per spin 
(a) for the $S=1/2$ Heisenberg model, and for CuGeO$_3$ 
(b) in the strong edge, and 
(c) in the weak edge. 
$\langle S^z_i\rangle^{\rm s}$ is used. 
The impurity density is proportional to $L^{-1}$. 
The width of the gaussian is $w=0.02$.
}
\label{fig:NMRL}
\end{figure}

Since eq.~(\ref{eq:chiHeis}) is the result for the infinite system, 
we consider finite-size effects.
In Figs.~\ref{fig:NMR0035}~I and II we show $g(\tilde H)$ for 
$\alpha=0$ and $0.35$ respectively, estimated by the data in 
Fig.~\ref{fig:Sz}~I and II. 
For $\beta\sim 1$ and regardless of the value of $\alpha$, 
$g(\tilde H)$ has the double-peak structure on account of the 
delocalization of $\langle S^z_i\rangle$ with mild curvature. 
Indeed, such a structure has been observed in the NMR spectrum 
for Sr$_2$CuO$_3$ at a low temperature [see Fig.~2 (30K) of 
ref.\ 78]. 
In this spectrum the interval between the peaks looks narrower 
than that in Fig.~\ref{fig:NMR0035}~I ($L=200$). 
This is interpreted as follows.
The amplitude of $\langle S^z_i\rangle^{\rm s}$ (and also 
$\langle S^z_i\rangle^{\rm u}$) becomes smaller and flatter 
with increasing $L$, as explained in \S5.2. 
Thus, the interval becomes narrower with increasing $L$, as 
observed in Fig.\ref{fig:NMRL}(a). 
Ultimately, they are combined into a single peak for $L\to\infty$. 
Experimentally, the chain length of Sr$_2$CuO$_3$ is estimated at 
$L=1800\pm500$, namely longer but finite. 
\par
\begin{fullfigure}
\figureheight{5cm}
\begin{center}
\end{center}
\caption{
Intensity of NMR spectra $g(\tilde H)$ for $\alpha=0$ (I) 
and $0.35$ (II) for the segment of $L=200$. 
The strong- and weak-edge cases are shown in (a) and (b), 
respectively. 
In II(a) the line shapes are almost the same for $\beta=0.40$-$0.95$. 
They are obtained from the {\it staggered part} of the data
in Fig.~\ref{fig:Sz}, since $\langle S^z_i\rangle^{\rm u}$ only 
weakly depends on $i$ for $\beta\sim 1$. 
The width of the gaussian is $w=0.02$.
}
\label{fig:NMR0035}
\end{fullfigure}
When the dimerization increases, the double peaks merge into 
a single peak, irrespective of the edge condition. 
This single peak is attributed to the disappearance of delocalized 
staggered moment, because the intensity of the peak at $\tilde{H}\sim 0$ 
originates from the dense distribution of 
$\langle S^{z}_{i}\rangle^{\rm s}$ near zero (see Fig.~\ref{fig:Sz}). 
There is one remarkable difference between the strong and weak 
edges; a broad skirt appears around the single peak for the latter,
as seen in Fig.~\ref{fig:NMR0035}(b).
This is caused by the localized free spins with large amplitude. 
\par

Finally, we refer to the influence of the impurity concentration. 
For $\beta=1$ the similar aspects to the Heisenberg case [see 
Fig.~\ref{fig:NMRL}(a)] are observed for arbitrary values of $\alpha$. 
In Fig.~\ref{fig:NMRL}(b) we show the system-size dependence of 
$g(\tilde H)$ for the spin-Peierls phase of CuGeO$_3$. 
With increasing $x$ ($1/L$) the height of the single peak lowers 
and the spectrum becomes broad. 
Such impurity effect seems common to various systems.
\par
%
\section{Haldane phase}
In this section, we study the edge effect in the Haldane phase,
and clarify the difference from the spin liquid phase. 
As a typical compound, we refer to a F-AF bond-alternating 
compound monoclinic CuNb$_2$O$_6$. 
As discussed in Appendix D, this material is well described by 
the model eq.~(\ref{eq:hamil}) with $\alpha=0$ and $\beta=-2.36$ 
($-2$) for the pure (10\%-doped) sample. 
To avoid confusion, we make sure of the terms for edge condition: 
Corresponding to the strong (weak) edge in the spin-Peierls regime, 
we use terms ``AF (F) edge" for $\beta<0$, and ``$J$ ($\beta J$) edge" 
for a general value of $\beta$. 
\par

\begin{figure}
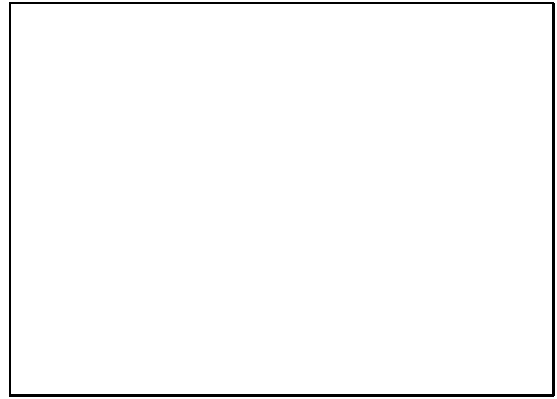

\figureheight{5cm}
\begin{center}
\end{center}
\caption{
Energy difference between the $J$ and $\beta J$ edges in the 
ground states. 
To show little size dependence, some systems of even $L$ are 
plotted together. 
$E(J)$ [$E(\beta J)$] indicates the total energy for the system 
of the $J$ ($\beta J$) edges. 
The minimum is $\Delta E/J=-0.75$ at $\beta=0$. 
The arrow indicates the reversal point at $\beta\sim -3$. 
This value 3 is roughly the ratio of energy gain in singlet 
($3J/4$) to triplet ($J/4$). 
}
\label{fig:edgeE}
\end{figure}

First of all, we consider the stability of the edge. 
In Fig.~\ref{fig:edgeE} we plot the energy difference $\Delta E$ 
in the ground states for a wide range of $\beta$ and $\alpha=0$. 
As mentioned for the spin-Peierls regime, the strong edge becomes 
stable, as soon as $\beta$ decreases from 1. 
However, as $\beta$ further decreases beyond $0$, $\Delta E$ starts 
to increase, and at last $E(\beta J)$ becomes lower than $E(J)$ 
at $\beta\sim -3$. 
For $\beta\to\infty$, $\Delta E$ becomes $|\beta|J/4=\infty$, 
because the number of F bonds is larger by one in the F-edge case. 
Thus, in the $S=1$ Haldane systems the F (weak) edge is always 
realized. 
Incidentally, the whole feature of $\Delta E$ including the 
reversal point $-3$ only weakly depends on the value of $\alpha$. 
In the bond-alternating systems with $|\beta|\ll\infty$ like 
monoclinic CuNb$_2$O$_6$, in which the lattice structure is 
rigid in contrast to the spin-Peierls compounds, two kinds of 
edge $J$ and $\beta J$ should be yielded at random by 
non-magnetic impurities. 
\par
%
Next, we discuss the spin polarization $\langle S^z_i\rangle$. 
As is known for the $S=1$ Heisenberg model,\cite{Kennedy,MiyashitaY} 
the magnetic excitation is localized at both edges as $S=1/2$ 
free spins. 
This localized moment decays exponentially with the localization 
length $\xi_{\rm edge}$, which is independent of the system size 
and is equivalent to the bulk correlation length 
$\xi_{\rm H}\sim 6.03$.\cite{MiyashitaY,WhiteH} 
Magnetic LRO's hardly appear even by doping, unless the interchain 
couplings are sufficiently strong.\cite{Ajiro,Kojima,Sakai}
Besides, localized edge spins have been actually observed by 
various experimental studies.\cite{Hagiwara} 
\par

\begin{figure}
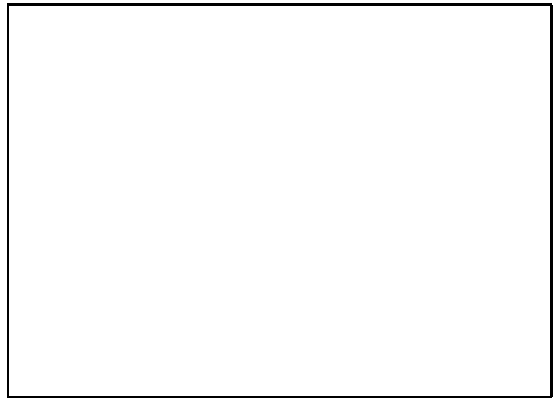

\figureheight{5cm}
\begin{center}
\end{center}
\caption{Comparison of spin polarization in the lowest magnetic
states between CuGeO$_3$ (strong edge) and CuNb$_2$O$_6$ (two 
kinds of edges) for the systems of (a) $L=100$ and (b) 50. 
As for parameter values, ($\alpha$,$\beta$)=(0.35,0.98) is assumed 
for CuGeO$_3$, and (0.0,-2.36) for CuNb$_2$O$_6$. 
}
\label{fig:FAFSz}
\end{figure}

We have already seen that such localization appears also in the 
present model, when the dimerization is introduced for the weak 
edge (Fig.~\ref{fig:Sz}). 
Then, we start with the F (weak) edge. 
In Fig.~\ref{fig:FAFSz} we compare $\langle S^z_i\rangle$ in 
the Haldane phase (CuNb$_2$O$_6$) with the spin liquid (CuGeO$_3$). 
For CuNb$_2$O$_6$ (open circle), $|\langle S^z_i\rangle|$ obeys 
typical exponential decay, in contrast to the spin-Peierls cases 
[see Fig.~\ref{fig:Sz}(b)]. 
Besides, $\xi_{\rm edge}$ is fairly short and is almost the same 
with the bulk correlation length $\xi \sim 3.3$ (Fig.~\ref{fig:crr2}). 
By comparing Figs.~\ref{fig:FAFSz}(a) and (b), we find that 
the amplitude of $\langle S^z_i\rangle$ near the edge is independent 
of the system size. 
On the other hand, not only the oscillating part but also the 
uniform part of $\langle S^z_i\rangle$ vanishes for the bulk. 
Since the situation $\langle S^z_i\rangle=0$ is the same with the 
ground state ($S=0$), the magnetic excitation never affects the bulk. 
Thus, the behaviors of the Haldane phase in the $S=1$ systems 
stays unchanged for the present case. 
Incidentally, the odd-site systems show similar behaviors to 
this case, as discussed in \S5.2.
\par
%
As for the AF (strong) edge, the excess spins extend to the 
whole system (small open diamond in Fig.~\ref{fig:FAFSz}). 
Nevertheless, the oscillating part is completely lacking. 
Thus, the magnetic excitations are not likely to induce LRO 
also in this case. 
Indeed, monoclinic $\rm CuNb_{2}O_{6}$ keeps the gap properties 
in heavily doped samples,\cite{Nishikawa} in spite of the 
relatively large interchain coupling (cf.~Appendix D). 
In contrast, for CuGeO$_3$ the excited moment extends to the 
whole system (solid circle in Fig.~\ref{fig:FAFSz}). 
Moreover, the amplitude of staggered part abruptly grows with 
decreasing $L$ (increasing $x$). 
Thus, the magnetic excitation leads to AFLRO, as has been 
already discussed. 
\par

\begin{figure}
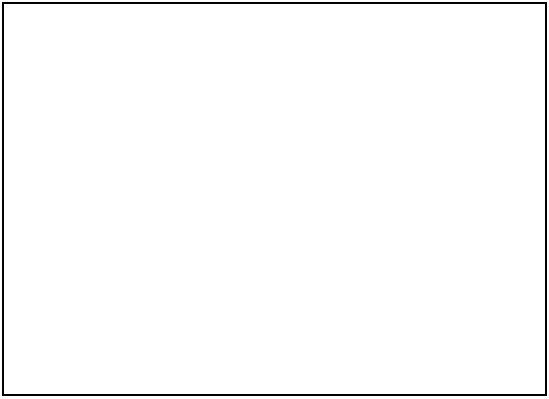

\figureheight{5cm}
\begin{center}
\end{center}
\caption{
System-size dependence of the NMR intensity per spin for CuNb$_2$O$_6$ 
(a) in the AF edges and (b) in the F edges. 
For odd-site systems, $g(\tilde H)$ is analogous to (b) with the 
doubled system size. 
We use the fixed value of $\beta=-2.36$ (for pure material) for 
simplicity, although the value of $\beta$ changes with the impurity 
concentration (see Appendix D). 
To obtain the spectra by eq.~(\ref{eq:NMR}), total $\langle S^z_i\rangle$ 
(including uniform part) is used with $w=0.02$. 
}
\label{fig:FAFNMR}
\end{figure}

Let us turn to the NMR spectra. 
In Fig.~\ref{fig:FAFNMR} we show the size dependence of the NMR 
intensity per spin for the parameters of pure $\rm CuNb_{2}O_{6}$. 
For a low impurity concentration (large $L$) $g(\tilde H)$ has a 
sharp peak at $\tilde H\sim 0$ for every edge condition, 
corresponding to the absence of large oscillating part of 
$\langle S^z_i\rangle$ in the bulk. 
For the F edge there appear skirts and broad side peaks due to 
the localized spins at the edges. 
Thus, the localization survives even in the low-temperature limit, 
so that the NMR line shape will hardly broaden with decreasing 
temperature, 
unlike the $S=1/2$ Heisenberg compound\cite{Takigawa} and the two-leg 
ladder system.\cite{Fujiwara} 
On the other hand, when the impurity concentration increases ($L$ 
becomes small), the height of the single peak at $\tilde H\sim 0$ 
lowers regardless of the edge condition. 
Simultaneously, the peak shifts slightly for the AF edge, while the 
intensity of the skirts and side peaks grow for the F edge. 
Anyway, the Haldane phase and the spin liquid share the tendency
toward the breakdown of the sharp peak by a few percent of impurity. 
\par
%
In conclusion, we recapitulate the relationship between the phase 
transition and the impurity effect. 
The characteristics of magnetic excitation in the Haldane phase 
are that the excited moment is confined in a finite range near 
the $\beta J$ edges, and that the oscillating part in 
$\langle S^z_i\rangle$ is almost absent for the $J$ edges. 
The Haldane phase occupies the whole $\alpha$-$\beta$ plane except 
for the $\alpha$ axis and $\alpha=\infty$, on which the phase is 
the spin liquid characterized by the delocalization of 
$\langle S^z_i\rangle$. 
As the parameters approach the spin liquid, the range of localized 
edge spins extends toward the bulk (independent of $L$) [$\beta J$ edge], 
and the amplitude of oscillation ($<1/L$) in $\langle S^z_i\rangle$ 
grows [$J$ edge]. 
In the thermodynamic limit (for the pure material), however, both 
the range of edge spins [$\beta J$ edge] and the amplitude of 
staggered moment [$J$ edge] are still negligible. 
On the other hand, when the chains are cut into finite segments 
(for the doped case), the range of edge spins may extend to the 
greater part of (or whole) segment [$\beta J$ edge], and the 
amplitude of staggered moment may becomes sufficiently large 
[$J$ edge], according to the balance between $\beta$ and $L$ ($x$). 
Thus, the interchain coupling, which is quantitatively important, 
has a chance to induce AFLRO. 
In the transition areas from the Haldane phase to the spin 
liquid, impurities play a dramatic role. 
\par

So far, we have not mentioned the two-leg ladder systems. 
In the typical ladder compounds, AFLRO is observed for the 
pure (LaCuO$_{2.5}$,\cite{Hiroi} $T_{\rm N}\sim 110$K) and 
lightly-doped (SrCu$_2$O$_3$\cite{Azuma,Fujiwara}) cases.
At the early stage these compounds were considered as isotropic 
ladders ($J_{\rm leg}/J_{\rm rung}=\alpha=1$), while recent 
experiments have concluded $\alpha\gsim 2$.\cite{ladder,Fujiwara} 
Actually, as described in \S3, the isotropic ladder point 
[$(\alpha,\beta)=(1,0)$] exhibits a typical properties of 
the Haldane phase with $\xi\sim 3$ without the tendency toward 
AFLRO. 
However, our preliminary calculations of various quantities 
as well as a previous study\cite{Laukamp} indicate that 
the aspects for $\alpha\gsim 2$ are rather similar to the 
$S=1/2$ Heisenberg model. 
Thus, it is probable that the impurity effect of SrCu$_2$O$_3$ 
will be similarly understood by considering that the system 
is situated in the transition area from the Haldane phase 
to the spin liquid in the two-chain Heisenberg limit ($\alpha\to\infty$). 
We will publish the results for ladder elsewhere. 
\par

%
\section{Summary} 
We have studied a variety of properties of the 1D Heisenberg model 
with the bond alternation and the NNN exchange coupling, using the 
DMRG and ED methods. 
We summarize our main results below. 
As for the bulk properties before doping: 
\par

(1)
The conditions for the Haldane phase are satisfied on the 
Shastry-Sutherland line ($\beta=2\alpha$) for $0\le\alpha<0.5$. 
A transition from the Haldane phase to a kind of spin liquid 
occurs at the Majumdar-Ghosh point ($\alpha=0.5$). 
\par

(2)
The above result can be extended to the whole $\alpha$-$\beta$ 
space. 
The Haldane phase continuously connects the $S=1$ limit 
($\beta=-\infty$) to the spin-Peierls regime ($\beta\sim 1$) 
through the bond-alternating regime, and the transition occurs 
at $\beta=1$. 
In the Haldane phase, there is no anomaly in the dimer and string 
order parameters, whereas the spin gap and the correlation length 
exhibit singularities due to the incommensurate correlation. 

(3)
We have determined the phase diagram of the incommensurate 
correlation. 
The range of essential incommensurate correlation defined by 
$\langle S_{i}^{z}S_{j}^{z} \rangle$ is quite different from 
that by $S(q)$. 
The singular points of the spin gap are included in the 
incommensurate range of $S_{ij}$. 

(4)
We have discussed various properties, especially dynamics, of 
monoclinic $\rm CuNb_{2}O_{6}$.
This compound is well described by eq.~(\ref{eq:hamil}) as a 
typical ferromagnetic-antiferromagnetic bond-alternating chain. 
\par

As for the edge effects, corresponding to the doping of non-magnetic 
impurities: 
\par

(5)
There are two kinds of chain ends, according as the edge bond 
is strong ($J$) or weak ($\beta J$). 
The strong edge is more stable for the spin-Peierls regime,
while the weak (F) edge becomes stable for $\beta\lsim-3$. 
This fact is possibly connected to the compositional phase transition 
of the first order between the dimerized and uniform AF phases 
in CuGeO$_3$. 
\par

(6)
The spin polarization is delocalized in the spin-Peierls systems 
and the two-leg ladders with large $J_{\rm leg}/J_{\rm rung}$, 
while the spin polarization is localized at the edge for the 
bond-alternating systems as well as the Haldane systems and the 
isolated dimers. 
In the former, $\langle S^z_i\rangle$ is enhanced by the impurity 
and triggers a 3D magnetic long-range order. 
In the latter, $\langle S^z_i\rangle$ is insensitive to the impurity, 
so that the gap properties are preserved. 
\par

(7)
For finite but large systems NMR spectra in the low temperatures 
show a double-peak structure in the vicinity of $\beta=1$ due 
to the delocalized spin polarization. 
On the other hand, a single sharp peak appears for the Haldane 
phase, regardless of the edge condition. 
In this phase, the conspicuous broadening of the spectra with 
decreasing temperature will not be observed. 
Generally, as the impurity concentration is increased, the height 
of peak lowers and eventually vanishes. 
\par

\section*{Acknowledgments}
We thank Masatoshi Sato and his group for providing us of 
experimental data on CuNb$_2$O$_6$ prior to publication as well as 
useful discussions.
We are grateful to Takashi Yamamoto, Kazuma Hirota and Zenji Hiroi 
for valuable discussions, and to Tota Nakamura for informing us
of some references.
A part of the numerical calculations was carried out with VPP500
at the Supercomputer Center of ISSP, University of Tokyo.
Parallelized fortran code was executed by SX4R at the Supercomputer
Center, Tohoku University and CP-PACS computer of Center for
Computational Physics, University of Tsukuba.
This study is partly supported by Grant-in-Aids for Encouragement
of Young Scientists given by the Ministry of Education, Science,
Sports and Culture.
\par

\appendix
\section{Relations between two representations}\label{chap:rep}
In this Appendix, we summarize the relations between the
Hamiltonians eq.~(\ref{eq:hamil}) and eq.~(\ref{eq:hamil2}).
Comparing corresponding terms, one obtains the relations,
\begin{equation}
\delta = \frac{1-\beta}{1+\beta}, \quad
{\tilde\alpha} = \frac{2\alpha}{1+\beta}, \quad
{\tilde{J}} = \frac{1+\beta}{2}J.
\end{equation}
Thereby, a line (point) in the $\alpha$-$\beta$ plane is
moved to another line (point) in the $\tilde\alpha$-$\delta$ plane.
Note that the Shastry-Sutherland line is $\beta=2\alpha$ in
eq.~(\ref{eq:hamil}) and $2\tilde\alpha+\delta=1$ in eq.~(\ref{eq:hamil2}). 
The isotropic ladder point is $(\alpha,\beta)=(1,0)$ or
$(\tilde\alpha,\delta)=(2,1)$.
However, the $\alpha$ axis ($\alpha$,$\beta=1$) remains the
$\tilde\alpha$ axis ($\tilde\alpha$,$\delta=0$).
The energy scale also should be changed;
$E$ in eq.~(\ref{eq:hamil}) is related to $\tilde{E}$
in eq.~(\ref{eq:hamil2}) as,
\begin{equation}
\frac{\tilde{E}}{\tilde{J}} = \frac{2}{1+\beta}\frac{E}{J}.
\end{equation}
\par

\appendix
\section{Cusp Behavior of the Gap}\label{chap:A-B}
In this Appendix we pursue the cusp behavior in $\Delta$ to 
supplement the discussions in \S3.1. 
\par

The second-order perturbation energies for the ground state 
($S=0$) and the lowest excited branches ($S=1$) around the 
isolated dimer point are given by eqs.~(\ref{eq:Esdimer}) 
and (\ref{eq:Etdimer}), respectively. 
The singlet energy is lowered by the second-order perturbation 
for $\beta\ne2\alpha$; but this decrement vanishes for 
$\beta=2\alpha$ (the SS line). 
This cancellation is understood intuitively as follows:
Two $\alpha$ bonds and one $\beta$ bond connect two singlet 
dimers, but $\alpha$ and $\beta$ play mutually inverse roles 
for the spin directions in the two singlet pairs. 
On the other hand, the triplet branch has a dispersion due to
the motion in the array of singlets for $\beta\ne 2\alpha$.
The bottom of $E_t(q)$ is located at $q=0$ for $\beta>2\alpha$ 
and at $q=\pi/2$ for $\beta<2\alpha$; the IC value of $q$ does 
not appear around the isolated dimer point. 
Thus, $\Delta$ is obtained as: 
\[ \frac{\Delta}{J} = \left\{
  \begin{array}{ll}
\displaystyle   1-\frac{1}{2}\beta^{2}
   & \mbox{$\beta=2\alpha$} \\
\displaystyle   1-\frac{1}{2}(\beta-2\alpha)
                            \\
\hspace*{1.8cm} \displaystyle
-\frac{5}{16}\beta^{2}+\frac{3}{4}\alpha^{2}-\frac{3}{4}\beta\alpha
   & \mbox{$\beta>2\alpha$} \\
\displaystyle   1+\frac{1}{2}(\beta-2\alpha)
                            \\
\hspace*{1.8cm} \displaystyle
  +\frac{3}{16}\beta^{2}-\frac{5}{4}\alpha^{2}-\frac{3}{4}\beta\alpha
   & \mbox{$\beta<2\alpha$.} 
  \end{array}\right. \]
Comparing the first-order terms, one finds the maximum of 
$\Delta$ is at $\beta=2\alpha$.
The cusp behaviors including the curvature in both sides of 
$\beta=2\alpha$ are reproduced by the differential coefficients:
\[ \left\{
  \begin{array}{ll} \displaystyle
\frac{\partial\Delta}{\partial \beta}<0, \quad 
\frac{\partial^{2}\Delta}{\partial \beta^{2}}<0 \qquad\qquad
& \mbox{$\beta>2\alpha$ } \\ \displaystyle
\frac{\partial\Delta}{\partial \beta}>0, \quad 
\frac{\partial^{2}\Delta}{\partial \beta^{2}}>0 \qquad\qquad
& \mbox{$\beta<2\alpha$.} 
  \end{array}\right. \] 
With increasing $\alpha$, however, the cusp point deviates to 
the side of $\beta<2\alpha$ as seen in Fig~\ref{fig:abplane}. 
\par

\begin{figure}
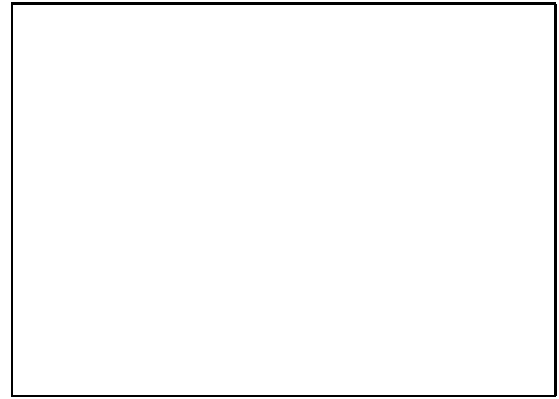

\figureheight{5cm}
\begin{center}
\end{center}
\caption{
Excitation energy of the lowest triplet branches for each 
available value of $q$ for $L=24$; (a) $\alpha=0.35$ and 
(b) $0.6$. 
The data are obtained by ED and the recursion method. 
}
\label{fig:cusp}
\end{figure}

\begin{figure}
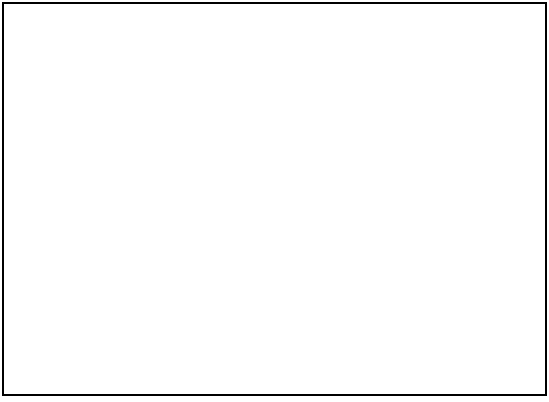

\figureheight{5cm}
\begin{center}
\end{center}
\caption{
Dynamical structure factor $S(q,\omega)$ for a cusp point 
($\alpha,\beta$)=(0.35,0.525). 
Intensity is proportional to the area of circle. 
Data for $L=14$-$26$ are plotted together. 
Note that every branch in this figure is independent of $L$.
}
\label{fig:flatdisp}
\end{figure}

Then, let us look at the cases of finite $\alpha$ with the ED 
results.
In Fig.~\ref{fig:cusp}(a) the lowest triplet excitation for
the available momenta in $L=24$ are plotted for $\alpha=0.35$.
The boundary between the lowest excitations of $q=\pi$ and
$q=\pi/2$ still forms the cusp near $\beta=0.525$.
An interesting point is that the excitation energies for
all the momenta meet at the cusp just like the isolated dimers.
No dispersion in the lowest excitation is evident in
$S(q,\omega)$ (Fig.~\ref{fig:flatdisp}); the excitation has
a local character at the cusp points.
\par

The above feature continues to $\alpha\sim 0.5$, beyond which the 
lowest excitation is no longer restricted to the C wave numbers.
As in Fig.~\ref{fig:cusp}(b) for $\alpha=0.6$, the wave number
of the lowest excitation does not switch directly from $q=\pi/2$
to $\pi$, but IC excitations lie between them.
Thus, the cusp behavior disappears for $\alpha\gsim 0.5$, as
seen for $\alpha=1$ in Fig.~\ref{fig:gapb}.
\par
%
\appendix
\section{Low-Energy States on the Shastry-Sutherland Line}\label{chap:SS}
The isolated dimer point satisfies the conditions of the Haldane 
phase itemized in \S2, while the MG point breaks the condition 
of a unique ground state. 
Hence, the phase is different between these points, which are 
the two ends of the SS line. 
With a phase transition in mind, we summarize and reconsider the 
knowledge of the ground state and low-lying excitations along 
the SS line in this Appendix. 
In \S C.1 using a non-local unitary transformation, we explain 
that the SS line strictly satisfies the conditions of the Haldane 
phase. 
In \S C.2 we summarize the structure of energy levels under 
different boundary conditions with the ED data. 
In \S C.3 we show that the transition occurs just at the MG point 
in terms of the perturbation theory for the degenerate case.
\par

\subsection{Non-local unitary transformation}
Following the formulation of the non-local unitary transformation
by Takada and Kubo,\cite{TakadaK,Takada} we rewrite the Hamiltonian
eq.~(\ref{eq:hamil}) as
\begin{eqnarray}
\label{eq:hnlu}
{\cal H}/J=&-&\sum^{L-1}_{i=1}
\bigl(\sigma^x_i\sigma^x_{i+1}+\tau^z_i\tau^z_{i+1}+
4\sigma^x_i\tau^z_i\sigma^x_{i+1}\tau^z_{i+1}\bigr)
\nonumber \\
&-&\alpha\sum^{L-1}_{i=1}
\Bigl[\sigma^x_i\tau^x_{i+1}+\sigma^z_i\tau^z_{i+1}
+\tau^x_i\sigma^x_{i+1}+\tau^z_i\sigma^z_{i+1}\Bigr.
\nonumber \\
\Bigl.&&\qquad\quad
+4\bigl(\sigma^x_i\tau^z_i\sigma^z_{i+1}\tau^x_{i+1}
+\tau^x_i\sigma^z_i\tau^z_{i+1}\sigma^x_{i+1}\bigr)\Bigr]
\nonumber \\
&+&\beta\sum^L_{i=1}{\mib \sigma}_i\cdot{\mib \tau}_i,
\end{eqnarray}
where the operator ${\mib S}$ is replaced by
${\mib \sigma}_i={\mib S}_{2i}$ and
${\mib \tau}_i={\mib S}_{2i-1}$.
As for the isolated dimer point ($\alpha=\beta=0$), the surviving
part [the first line of eq.~(\ref{eq:hnlu})] is coupled Ising chains
with a $Z_2\times Z_2$ symmetry.
The ground state has ferromagnetic orders in both $\sigma^x_i$
and $\tau^z_i$, and is fourfold degenerate ($\sigma_i^x=\pm 1/2$
and $\tau_i^z=\pm1/2$); the symmetry is completely
broken.\cite{KohmotoT,Takada}
Moreover, since the eigenstate is $\psi_1$ of eq.~(\ref{eq:psi1})
(or $\psi_2$), there exists a single-triplet gap of magnitude $J$.
The string order parameter is calculated as $1/4$. 
Thus, the conditions of the Haldane phase are satisfied at 
this point.
These conditions are still satisfied on the SS line for 
$0\le\alpha<0.5$, because the ground state is identical with 
that of the isolated dimer point, and is unique with a finite 
singlet-triplet gap.
We will make sure of the latter point below. 
This result has been also reached with the matrix product 
procedure.\cite{Brehmer1} 
\par

\subsection{Structure of low-energy levels}
Here, we study the low-energy levels on the SS line 
($\delta+2\tilde\alpha=1$) for finite systems. 
The boundary condition greatly affects the level structure.
\par

\begin{figure}
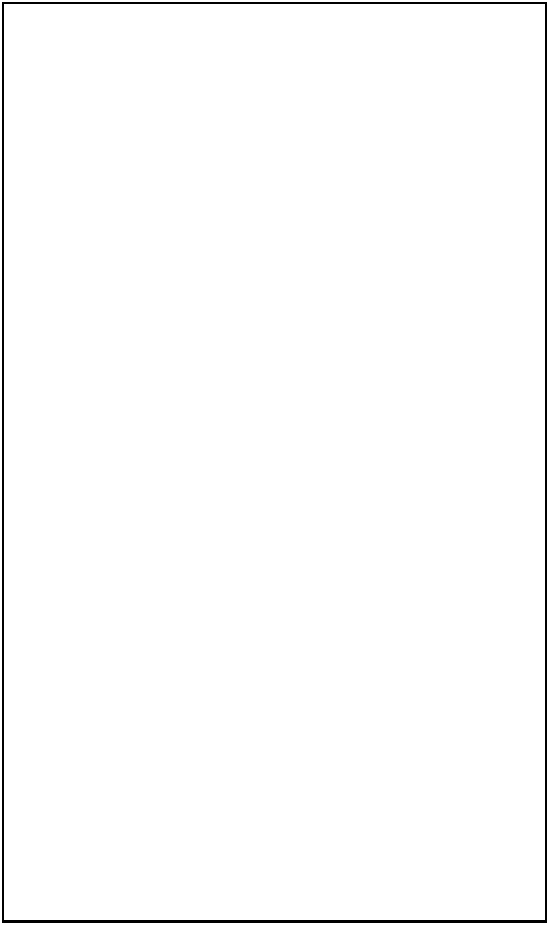

\figureheight{12cm}
\begin{center}
\end{center}
\caption{
Some lowest levels in the $S^z=0$ subspace on the SS line 
($\delta+2\tilde\alpha=1$) as a function of $\tilde\alpha$; 
(a) PBC ($L=16$),
(b) OBC with the strong edges ($L=12$), and
(c) OBC with the weak edges ($L=12$). 
The insets are respective magnifications near the MG point. 
The arrow in the inset of (a) indicates the crossing point of 
the second lowest $S=0$ level and the lowest $S=1$ level.
}
\label{fig:level}
\end{figure}

\begin{figure}
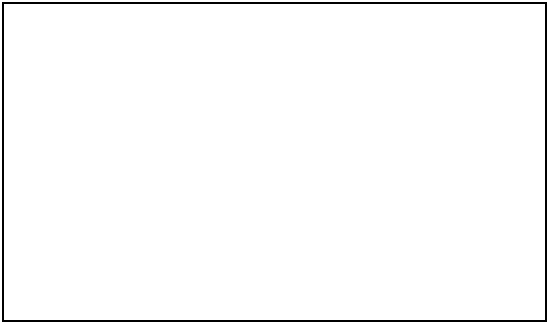

\figureheight{4cm}
\begin{center}
\end{center}
\caption{
Three lowest levels on the SS line ($\delta+2\tilde\alpha=1$) 
near the MG point in the $S^z=0$ sector under PBC. 
The system size is different between (a) and (b) [see also 
the inset of Fig.~\ref{fig:level}\ (a)]. 
The Arrow indicates the crossing point of the lowest $S=1$ 
level and the second lowest $S=0$ level. 
}
\label{fig:levelsize}
\end{figure}

Figure \ref{fig:level}(a) shows the ED result under PBC in the 
$S^z=0$ sector. 
Under this boundary condition there is no degeneracy in the 
ground state in accordance with the conditions of the Haldane phase. 
At the isolated dimer point, the first excited states are triplet
and $3L/2$-fold degenerate.
Although this degeneracy is lifted by a finite value of 
$\tilde\alpha$, the energy difference is very small except for 
the vicinity of the MG point.
There, as seen in the inset, the level of $S=0$, which is one of 
the degenerate ground states at the MG point, goes up linearly 
with decreasing $\tilde\alpha$, and intersects the lowest $S=1$ 
level at $\tilde\alpha\sim 0.48$ (indicated by arrow).
This means that the lowest excited state becomes singlet near the
MG point.
However, the range of this singlet-singlet excitation depends on
the system size.
In Fig.~\ref{fig:levelsize} we show similar figures with different
system sizes.
With increasing $L$, the crossing point approaches the MG point.
This point will be pursued in \S C.3.
\par

In Fig.~\ref{fig:level}(b) plotted is the result for OBC with
the strong edges. 
In this case the ground state is either $\psi_1$ or $\psi_2$
according as singlet pairs sit on the edge bonds, and always 
nondegenerate.
Note that the edge bonds are occupied by singlet pairs even 
for the non-dimerized (MG) case. 
The first excited state is always triplet. 
Thus, not only the ground state is identical with that of PBC, 
but properties of excitation are also analogous to those of PBC.
We have taken advantage of such a property to obtain the bulk 
gap in \S3.1. 
\par

Figure~\ref{fig:level}(c) shows the same for OBC with the weak 
edges. 
In this case, there is the pseudo fourfold (twofold in the $S^z=0$
sector) degeneracy except for the vicinity of the MG point. 
The ground state ($S=0$) is neither $\psi_1$ nor $\psi_2$ except 
for the MG point, due to the isolated spins at the edges. 
Thus, this edge condition has a feature specific to the open edge 
of the Haldane systems. 
The nearly degenerate lowest levels ($S=0$ and $S=1$) separate 
at $\tilde\alpha\sim 0.46$; the $S=0$ level abruptly comes down 
and converges to the value of $\psi_1$ (or $\psi_2$) at the MG point. 
In this process the singlet array is rearranged, say from L to R, 
to adapt itself to the strong edges. 
On the other hand, near the above separation point the second 
lowest $S=0$ level, which has a similar energy to the lowest 
$S=1$ level near the MG point, abruptly changes its direction 
and merges into the upper $S=1$ levels. 
These behaviors correspond to the level crossing under PBC mentioned 
above, and the separation point also highly depends on the system 
size. 
This point will be taken up again in \S C.3.
\par

In this connection, recent studies\cite{Hassan} have discussed 
the appearance of the mid-gap states for the doped CuGeO$_3$ at 
$T_{\rm N}<T<T_{\rm SP}$. 
According to the above results, however, mid-gap levels are not 
likely to arise from the strong edges, the case for CuGeO$_3$. 
On the other hand, for the weak edges (including odd-site systems) 
the energy difference among the nearly fourfold degenerate states 
increases exponentially with $1/L$. 
Thus, the experiment may detect such levels, if any.
\par

Finally, we generalize the level structure for the bond-alternating 
systems, aside from the behaviors near the MG point. 
Under PBC or OBC with the strong edges, the ground state is a 
nondegenerate singlet ($\psi_1$ or $\psi_2$ in the case of 
the SS line) with the singlet-triplet gap over it. 
Under OBC with the weak edges, the ground state is nearly fourfold 
degenerate (one singlet and three triplets) with the bulk gap 
over these four states. 
\par

\subsection{Breakdown of the perturbation theory}
Having seen extraordinary behaviors around the MG point 
in \S C.2, here we reconsider them in the light of the 
perturbation theory for the degenerate case, and shows that 
the non-Haldane phase is restricted to the MG point. 
\par

The Hamiltonian eq.~(\ref{eq:hamil2}) on the SS line is written
around the MG point as,
\begin{equation}
{\cal H}={\cal H}_{\rm MG}+\zeta {\cal H}',
\end{equation}
\begin{equation}
\label{eq:hamilS2}
{\cal H}_{\rm MG}=\tilde J\sum_{i=1}^L\left[{\mib S}_{i}\cdot{\mib S}_{i+1}
          +\frac{1}{2}{\mib S}_{i}\cdot{\mib S}_{i+2}\right],
\end{equation}
\begin{equation}
\label{eq:hamilS2}
{\cal H}'=-2\tilde J\sum_{i=1}^L\left[(-1)^i{\mib S}_{i}\cdot{\mib S}_{i+1}
          +\frac{1}{2}{\mib S}_{i}\cdot{\mib S}_{i+1}\right],
\end{equation}
where we assume PBC and $\zeta=1/2-\tilde\alpha$ to be infinitesimal.
Here, ${\cal H}_{\rm MG}\psi_n=E_{\rm MG}\psi_n$ ($n=1,2$) holds,
where $E_{\rm MG}=-3L\tilde J/8$ is the ground-state energy at the MG point,
and $\psi_n$ is assumed to be normalized.
Let $\psi_1$ be the ground state on the SS line:
${\cal H}'\psi_1=\varepsilon_1\psi_1$ with
$\varepsilon_1=-3L\tilde J/4$.\cite{SS}
Since $\psi_1$ and $\psi_2$ are orthogonal in the thermodynamic
limit and $\psi_1$ is still an eigenstate under the perturbed
Hamiltonian, the first-order perturbation energies for the
two levels are formally written in a simple form, as
\begin{equation}
E'_n=\zeta\langle\psi_n|{\cal H}'|\psi_n\rangle=\zeta\varepsilon_n
\qquad (n=1,2).
\end{equation}
Thus, the perturbation theory naturally works for the lower level
$\psi_1'=\psi_1$.
As for the higher level $\psi_2'$, if the perturbation theory
works, the energy is also proportional to $1/2-\tilde\alpha$
near the MG point, actually as in the inset of Fig.~\ref{fig:level}(a)
and in Fig.\ref{fig:levelsize}.
Since a direct calculation of $\varepsilon_2$ looks difficult,
we estimate it by ED with PBC through,
\begin{equation}
\varepsilon_2=\langle\psi_2|{\cal H}'|\psi_2\rangle=
\left.\frac{\partial E_2}{\partial\zeta}\right|_{\zeta\to 0},
\end{equation}
where $E_2$ is the total energy with respect to $\psi_2'$.
\par

\begin{figure}
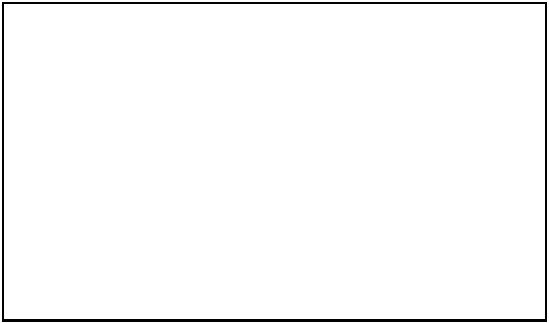

\figureheight{4cm}
\begin{center}
\end{center}
\caption{
(a) System-size dependence of the inverse of the first-order
perturbation energy per site for $\psi_2'$ in PBC.
The corresponding value for the lowest level in OBC with the
weak edges are also shown.
(b) System-size dependence of the crossing point of $\psi_2'$
and the lowest $S=1$ level in PBC.
}
\label{fig:extrap}
\end{figure}

In Fig.~\ref{fig:extrap}(a) we plot $\tilde J/\varepsilon_2$ thus
obtained versus $1/L$.
It shows that $\varepsilon_2$ diverges in the thermodynamic limit,
namely the perturbation breaks down for $\psi_2'$.
Along with this phenomenon, the crossing point of $\psi_2'$ and
the lowest $S=1$ level converges to the MG point [Fig.~\ref{fig:extrap}(b)].
Thus, one of the degenerate levels at the MG point never continues 
to the dimerized region ($\delta>0$); the first excited state
under PBC is always triplet on the SS line.
\par

On the other hand, as pointed out for OBC with weak edge in \S C.2,
the lowest level, which separates from the triplet levels near the
MG point, behaves similarly with $\psi_2'$ in PBC.
We also estimate the perturbation energy $\varepsilon$ for this case
in the same manner, and plot $\tilde J/\varepsilon$ together in
Fig.~\ref{fig:extrap}(a).
Within the numerical accuracy, the values of $\tilde J/\varepsilon_2$
and $\tilde J/\varepsilon$ are identical.
Thus, the lowest level ($S=0$) at the MG point does not smoothly 
continue to the $S=0$ state which makes up the fourfold degeneracy 
for $\delta>0$.
And the degeneracy of the ground state continues to the limit
$\tilde\alpha\to 1/2$.
Thus, the energy difference between the strong and the weak edges 
in the limit of $L\to\infty$ and $\tilde\alpha\to 1/2$ becomes 
the MG gap. 

\par

From the above we conclude that the Haldane phase continues to this
limit, and that the MG point is critical.
Hence, the consequence of \S C.1 is valid for $\delta\ne 0$. 
On the other hand, we can say inversely that in finite systems 
the properties of the Haldane phase are lost near the MG point. 
So far, we have restricted our discussion to the SS line.
As far as we have tried numerically, however, this discussion is 
applicable to any path which starts from an arbitrary point 
on the $\alpha$ axis ($\alpha>\alpha_{\rm c}$) and advances in 
an arbitrary direction of finite $\delta$.
\par

\section{Bond-alternating compound \\
CuNb$_2$O$_6$}\label{chap:CuNb2O6}
In this Appendix we discuss various aspects of a F-AF 
bond-alternating chain monoclinic CuNb$_2$O$_6$ to supplement 
the background of \S6. 
The dynamical properties of bond-alternating chain, in particular
for $\beta<0$, are surveyed in \S D.1. 
In \S D.2 various properties of CuNb$_2$O$_6$ are summarized, 
before discussing its dynamical properties in \S D.3. 
\par

\subsection{Spin dynamics for $\alpha=0$}
Since principal static properties have been already described 
in this paper as well as in some studies,\cite{Hida,Takada,Borras}
we pay attention to the dynamics. 
As discussed in \S5.1, low-lying spin excitation change its 
character from even-number spinons to magnons, as soon as the 
bond alternation is introduced. 
Consequently, isolated magnon branches appear below the continuum.
As in Figs.~\ref{fig:sqw0}(a) and (b), the ratio of intensity 
(residue) in the lowest isolated branch increases with decreasing 
$\beta$ from 1. 
At the isolated dimer point this branch has the total intensity 
with a completely flat band at $\omega=J$. 
For $\beta>0$ the wave number of gap is $\pi$ and the band top 
is at $\pi/2$. 
\par

\begin{fullfigure}
\begin{center}
\figureheight{7cm}
\end{center}
\caption{
Dynamical structure factor $S(q,\omega)$ for four values of $\beta$ 
($\delta$) and $\alpha=0$.
Data obtained by the ED with a recursion method for $L=14$-$26$ 
are simultaneously plotted.
The intensity is proportional to the area of circle. 
In (a) and (b), to convert the $\tilde\alpha$-$\delta$ 
representation into the $\alpha$-$\beta$ representation, 
5/6 and 2/3 have to be multiplied to $\omega$, respectively.
See Appendix~A.
}
\label{fig:sqw0}
\end{fullfigure}
%
On the other hand, as $\beta$ decreases further for $\beta<0$, 
the main branch comes to have dispersion with a minimum at 
$q=\pi/2$, as discussed by the perturbation calculations in 
Appendix~B. 
A continuum due to two triplet pairs appears around $\omega=2J$, 
as pointed out for $\beta\sim 0$ by Hida.\cite{Hida} 
Furthermore, another isolated branch appears approximately at 
$\omega=|\beta|J$, which corresponds to the energy to break 
one F bond. 
In addition, higher harmonics appear, \eg at $\omega\sim 6.5J$ 
in Fig.~\ref{fig:sqw0}\ (d).
\par

\begin{figure}
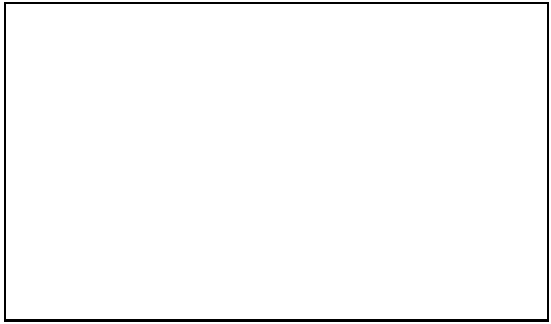

\figureheight{4cm}
\begin{center}
\end{center}
\caption{
(a) Intensity [$S(q,\omega)$] of LEB at $q=\pi/2$ (solid circle) 
and of HEB at $q=\pi$ (open circle) as a function of $-\beta$.
The data points are extrapolated from the ED results of
$L=14$-$26$.
(b) Excitation energy of LEB at $q=\pi/2$ [$\omega_{\rm L}(=\Delta)$] 
and of HEB at $q=\pi$ [$\omega_{\rm H}$], and their ratio are 
shown as a function of $-\beta$.
These values are estimated by the DMRG and ED methods ($L\to\infty$)
for eq.~(\ref{eq:hamil}) with $\alpha=0$.
The experimental value of the ratio for 10\%-doped CuNb$_2$O$_6$ are
indicated by the dashed line.
}
\label{fig:wvsb}
\end{figure}

To begin with, let us look at the dominant lowest-energy branch 
(LEB) and the continuum. 
As shown in Fig.~\ref{fig:Sq}, the maximum of integrated 
intensity $S(q)$ moves from $q=\pi$ to $\pi/2$ through IC wave 
numbers, as $\beta(<0)$ decreases.
According to this change, and because the greater part of 
intensity is concentrated on LEB, the weight in LEB moves 
from $q=\pi$ to $\pi/2$. 
For $\beta\lsim -3$ the intensity at $q\sim\pi/2$ is the 
strongest, and grows further with decreasing $\beta(<0)$ 
[Fig.\ref{fig:wvsb}(a)]. 
Similarly, the weight of the continuum chiefly distributed 
at $q\sim\pi$ for small values of $|\beta|$ gradually moves 
to $q\sim\pi/2$.
We have found for $\beta\lsim -2$ that the residue of LEB 
for $0.2<q/\pi<0.5$ is more than 85\% and scarcely depend 
on the system size, whereas for $q\lsim 0.2\pi$ it considerably 
decreases with increasing $L$. 
This implies that LEB enters the continuum at $q\sim 0.2\pi$.
A similar aspect was observed for the $S=1$ Heisenberg 
model.\cite{Takahashi}
\par
%
Next, we mention the high-energy branch (HEB) near
$\omega\sim |\beta|J$.
The dispersion of HEB is flatter with a minimum also at $q/2$.
Accordingly, local destruction of a ferromagnetic bond chiefly
contributes to this excitation.
There exist weak continuum spectra around HEB, although HEB
itself is an isolated branch because of little dependence of
the intensity on $L$.
As shown in Fig.~\ref{fig:wvsb}(a) the intensity of HEB increases
with decreasing $\beta$ at first; the residue for HEB at $q=\pi$ 
becomes as large as 70\% for $\beta<-4$ in spite of the high-energy 
process. 
With decreasing $\beta (<0)$ further, however, its intensity
decreases and vanishes in the Haldane limit, according to the
total intensity $S(\pi)$ (Fig.~\ref{fig:Sq}).
Therefore, this branch is specific to the F-AF bond-alternating
systems.
\par

\subsection{Monoclinic CuNb$_2$O$_6$}
As a spin-gap compound CuNb$_2$O$_6$ has been actively studied
by various experimental means
recently\cite{KodamaGap,Nishikawa,Fukamachi,Mitsuda,KodamaINS}.
CuNb$_2$O$_6$ has zigzag chains of edge-sharing CuO$_6$ octahedra
with localized $S=1/2$ spins on Cu$^{2+}$ ions.
This compound has two polymorphs, \ie orthorhombic and
monoclinic phases.
The orthorhombic phase is considered as a 1D Heisenberg antiferromagnet
with an AFLRO below 7.3K; on the other hand, the
monoclinic phase shows thermal-activation-type behaviors in
magnetic susceptibility $\chi(T)$.\cite{KodamaGap}
Henceforth, we restrict our discussions to the monoclinic phase.
Owing to the zigzag structure and the neutron scattering data
of the lowest excitation at $q=\pi/2$, it is expected that
monoclinic CuNb$_2$O$_6$ should be a F-AF bond-alternating chain.
Actually, the experimental $\chi(T)$ was fitted successfully
to the results of the model eq.~(\ref{eq:hamil}) with $J=51$K,
$\alpha=0$ and $\beta=-2.36$.\cite{Nishikawa}
Simultaneously, the coincidence in specific heat $C(T)$ is 
satisfactory with the same parameter set.
Our estimation $\Delta/J=0.3714$ for these values (Fig.~\ref{fig:gapb})
leads to $\Delta=19$K.
\par

Notwithstanding, recalling the ambiguity in modeling a spin-gap 
systems based solely upon $\chi(T)$, for example 
(VO)$_2$P$_2$O$_7$\cite{Garret} and KCuCl$_3$,\cite{NakaK} 
one should run a careful check by other means, in particular 
neutron scattering.
\par

Before discussing the neutron experiment, we touch on a couple 
of related properties of CuNb$_2$O$_6$. 
First, we consider the effect of the NNN coupling.
Fitting the experimental $\chi(T)$ to the ED results of
$L=16$\cite{note6-1} around $(\alpha,\beta)=(0,-2.36)$, we have 
found that a series of parameter sets such as 
$(0.10,-2.0)$, $(0.24,-1.5)$ and $(0.38,-1.0)$ bring favorable 
results in the same degree with $(0,-2.36)$.
This reminds us of the tendency that $\alpha$ encourages
the $\uparrow\uparrow\downarrow\downarrow$ correlation.
Concerning $C(T)$, however, such a set leads to a poorer fit, 
as $\alpha$ increases.
Hence, we conclude that the NNN exchange is negligibly small in
CuNb$_2$O$_6$ despite the edge-sharing linkage.
This is intuitively understood by considering the difference
in the orbital structure between CuGeO$_3$ and CuNb$_2$O$_6$.
In CuGeO$_3$ the NNN exchange arises from a superexchange process
through the path of Cu-O-O-Cu in the CuO$_2$ plane.\cite{Khomskii}
In this case a hybridized Cu-O orbital overlaps the counter
Cu-O orbital to some extent.
On the other hand, in CuNb$_2$O$_6$ the two Cu-O hybridized orbitals
of our concern are not in a plane and do not overlap each other
because of the zigzag linkage of CuO$_6$ octahedra.
This comparison affords vital support to the above mechanism
of the NNN exchange in the edge-sharing compounds.
\par

Next, we consider the effect of elemental substitution.\cite{Nishikawa}
In the crystal of monoclinic CuNb$_2$O$_6$, Cu can be replaced
by Zn up to 40\%.
In sharp contrast to CuGeO$_3$, the gap behaviors remain and no
magnetic LRO appears even for the 40\%-doped sample.
In doped samples, however, lattice constants and bond angles
change according to the Zn concentration, mainly due to the
different atomic radii between Cu and Zn; this change, particularly
in the bond angles, ought to affect the exchange integral,
according to the Kanamori-Goodenough rule.\cite{Kanamori}
Actually, considerable changes in parameters occur as summarized
in Table~\ref{table:param},\cite{noteSato} where the values
are estimated by the fit of $\chi(T)$ for some doped
samples Cu$_{1-x}$Zn$_x$Nb$_2$O$_6$.
Also entered in Table~\ref{table:param} is the gap estimated
through eq.~(\ref{eq:hamil}) using those parameters. 
The decreasing tendency versus $x$ coincides with the gap
estimated from $C(T)$, which decreases linearly with $x$.\cite{Nishikawa}
This behavior is contrary to what the $S=1$ Haldane systems 
show; the gap increases with increasing $x$.\cite{Kikuchi}
The reason is that thick doping cuts the chains into shorter 
segments, and that the gap generally increases as the system 
size becomes small.
In the case of CuNb$_2$O$_6$ the decrease of $|J|$, which is
specific to this compound, probably overcomes the system-size
dependence of the gap and the increase of $\beta$.
\par

\begin{table}
\caption{
Parameters for the doped CuNb$_2$O$_6$ obtained from the fit of
$\chi(T)$.
The gap $\Delta$ is estimated through the model
eq.~(\ref{eq:hamil}).
}
\label{table:param}
\begin{tabular}{@{\hspace{\tabcolsep}\extracolsep{\fill}}cccc} \hline
 $x$ & $J$\ (K) & $\beta$ & $\Delta$ (K) \\ \hline
 0.0 &     51.4 &   -2.36 & 19.1 \\
 0.1 &     44.5 &   -1.95 & 18.9 \\
 0.2 &     40.0 &   -1.86 & 17.6 \\
 0.3 &     33.4 &   -1.43 & 17.2 \\
 0.4 &     28.1 &   -1.03 & 17.1 \\ \hline
\end{tabular}
\end{table}
%
\subsection{Inelastic neutron scattering}
Recently an inelastic neutron scattering experiment was
performed on an Zn-doped CuNb$_2$O$_6$ sample.\cite{KodamaINS}
The spectra obtained in this experiment have reproduced the
most features of the F-AF bond-alternating chain discussed in
\S D.1.
We owe all the experimental data in this subsection to ref.~15.
\par

We begin with the evaluation of the parameters.
In Fig.~\ref{fig:wvsb}(b) shown are the excitation energies
of LEB $\omega_{\rm L}$ at $q=\pi/2$ ($\omega_{\rm L}=\Delta$), 
and of HEB $\omega_{\rm H}$ at $q=\pi$, and their ratio
$\omega_{\rm H}/\omega_{\rm L}$, estimated by
the DMRG and ED methods for eq.~(\ref{eq:hamil}) with $\alpha=0$.
On the other hand, the experimental values are
$\omega_{\rm L}^{\rm ex}=1.73$meV and
$\omega_{\rm H}^{\rm ex}=13.14$meV, respectively.
Then, the ratio becomes
$\omega_{\rm H}^{\rm ex}/\omega_{\rm L}^{\rm ex}=7.6$, which is
shown with dashed line in Fig.~\ref{fig:wvsb}(b).
Thus, we find that the value $\beta$ for CuNb$_2$O$_6$ of 10\% 
substitution is about $-2.0$, which agrees well with the
values estimated by $\chi(T)$ in Table~\ref{table:param}.

\begin{figure}
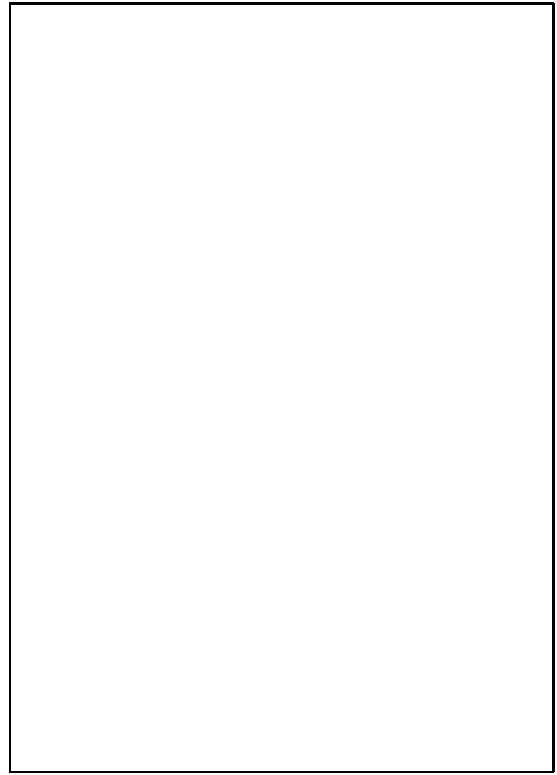

\figureheight{10cm}
\begin{center}
\end{center}
\caption{
Excitation energies (a) of LEB for some values of $\beta$, 
and (b) of HEB for a few values of $\beta$ ($\alpha=0$) and 
for some values of $\alpha$ ($\beta=-2.0$), as a function of $q$.
The data for $L=14$-$26$ are simultaneously plotted, which are 
obtained by ED and recursion methods. 
A large gray circle with a error bar represents the experimental 
data of 10\%-doped CuNb$_2$O$_6$\cite{KodamaINS} with $J=46.9$K. 
}
\label{fig:polelh}
\end{figure}

In Figs.~\ref{fig:polelh}(a) and (b), we depict the excitation
energy of LEB and of HEB, respectively, for the whole $q$ space.
In these figures we also plot the experimental results with 
$J=46.9$K, which best fits the whole data (both for LEB and HEB)
at the same time to the calculated values for $\beta=-2.0$.
They agree well as a whole except for $q=0.8\pi$ in LEB, 
where the experimental accuracy is rather poor due to the 
weak intensity.
The above value of $J$ is consistent with $J=47.9$K obtained 
from the gap $\Delta$.
In Fig.~\ref{fig:polelh}(b) we also show the data of $\alpha\ne 0$
and $\beta=-2.0$.
Although HEB is concave for $\alpha=0$ irrespective of the
value of $\beta$, as seen in Figs.~\ref{fig:sqw0}(c) and (d), 
yet it becomes convex as $\alpha$ increases.
This feature for finite $\alpha$ contradicts the experimental
results.
Thus, the neutron experiment supports the absence of the NNN 
exchange.
\par
\begin{figure}
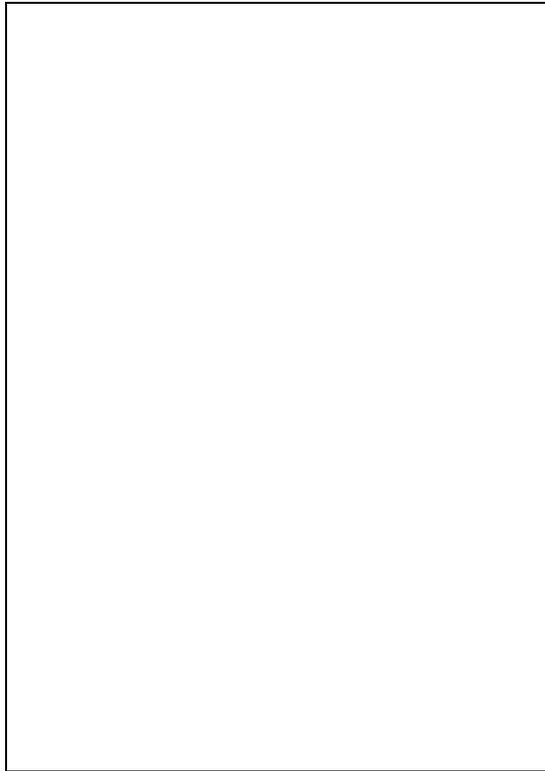

\figureheight{10cm}
\begin{center}
\end{center}
\caption{
Intensity [$S(q,\omega)$] of (a) LEB for several values of $\beta$, 
and of (b) HEB for a few values of $\beta$; both for $\alpha=0$ 
and as a function of $q$.
The data for $L=14$-$26$ are simultaneously plotted, obtained from 
the ED and recursion methods.
The diffuse data points for $\beta=-1.5$ in (b) is owing to the 
overlap with the continuum spectrum around HEB. 
A large gray circle with a error bar represents the experimental 
value of 10\%-doped CuNb$_2$O$_6$ scaled so as to best fit the 
lowest-energy branch of $\beta=-2.0$. 
}
\label{fig:residuelh}
\end{figure}

%
In Figs.~\ref{fig:residuelh}(a) and (b) shown are the intensity
[$S(q,\omega)$] of LEB and of HEB, respectively.
In these figures we scale the experimental results so that
LEB may be best fitted.
For LEB in Fig.~\ref{fig:residuelh}(a) the experimental intensity
is roughly reproduced by the calculated values for $\beta=-2.0$. 
Kodama \etal pointed out the conspicuous intensity at 
$q=0.6\pi$.\cite{KodamaINS} 
This is probably because the system ($\beta=-2.0$) is in the regime 
of IC correlation in $S(q)$ as discussed in \S4.2.
Then, $S(q)$ has a peak at $q=0.515\pi$ (See Fig.\ref{fig:qmax}).
On the other hand, the experimental intensity for HEB 
[Figs.~\ref{fig:residuelh}(b)] is considerably smaller than
what the theory expects.
This discordance is still an open question.

%


\vfil\eject
\vfil\eject

\begin{thebibliography}{99}
\bibitem{noteq} Since we discuss also the weight of $S(q,\omega)$, the 
original Brillouin zone, namely one for $S=1/2$ Heisenberg model, is 
used throughout this paper.
\bibitem{sP} When the $\delta$ dependence of the spin gap in 
eq.~(\ref{eq:hamil2}) is derived from the $1-\beta$ dependence in 
eq.(\ref{eq:hamil}), one should pay attention to the simultaneous 
change of $\alpha$ to $\tilde{\alpha}$. 
See Appendix~A.
\bibitem{note32} Indeed, there is an oscillation with a characteristic 
period in $S_{ij}$.
We take up this point in \S4.
\bibitem{Sznote} A part of the following aspects as to the local moment
have been already reported in recent literature.
See refs.\ 53-55.
\bibitem{note4spinon} For instance, four humps appear in the $S^z=-2$ 
subspace, as we expect four spinons, there.
\bibitem{notealpha} One of the present authors concluded in a previous
paper\cite{YS} that the value of $\tilde\alpha$ was $0.4$-$0.45$ by
comparing $S(q,\omega)$ with the inelastic neutron scattering spectra
for CuGeO$_3$ in ref.\ 65.
Afterward, however, more elaborate examination of neutron spectra
by Fujita and Arai has corrected the value to 0.3-0.4, which is
compatible with that obtained by spin susceptibility,
$\tilde\alpha\sim 0.35$.\cite{RD-Castilla,Klumper}
See ref.\ 66 for details.
\bibitem{note6-1} The difference between $L=16$ and $12$ for $\chi(T)$
                  and $C(T)$ is hardly observed for sufficiently
                  dimerized cases ($1\gg\beta\gg -\infty$).
\bibitem{noteSato} We owe the values in Table~\ref{table:param} to M.\ Sato's
                   group of Nagoya University. As $x$ increases, the
                   accuracy of fit naturally becomes poor.
\end{thebibliography}
\end{document}